\documentclass{article}
\usepackage{amssymb,amsmath}
\usepackage{latexsym}
\usepackage{graphicx}
\usepackage[dvips]{color}
\usepackage{epstopdf}
\usepackage{hyperref}
\newcommand{\la}{\frak{a}}
 \newcommand{\lN}{\frak{n}}
 
\def\be{\begin{equation}}
\def\ee{\end{equation}}
\def\beq{\begin{eqnarray}}
\def\eeq{\end{eqnarray}}
\def\bearl{\begin{array}{l}}
\def\bearll{\begin{array}{ll}}
\def\bearlll{\begin{array}{lll}}
\def\eear{\end{array}}
\def\beqe{\begin{equation}}
\def\eeqe{\end{equation}}

\newcommand{\LL}{\mathbb L}
\newcommand{\DD}{\mathbb D}
\renewcommand{\SS}{\mathbb S}

\newcommand{\R}{\mathbb R}
\newcommand{\MM}{\mathbb M}

\renewcommand{\gg}{\mathfrak g}
\renewcommand{\ss}{\mathfrak s}
\newcommand{\kk}{\mathfrak k}
\newcommand{\pp}{\mathfrak p}
\newcommand{\Der}{\mathfrak{Der}}

 \newcommand{\SL}{\mbox{SL}(2,\mathbb{R})}

\newcommand{\p}{\partial}

  \def\ts#1#2{\ \mbox {\kern0.1ex\raise1.6ex\hbox{\tiny
 #1}\kern-0.45em\mbox{$\star$}\kern-0.45em\lower1.2ex\hbox{\tiny
 #2}}}

\def\st#1#2{\ \mbox {\kern0.1ex\raise1.6ex\hbox{\tiny
#1}\kern-0.45em\mbox{$\star$}\kern-0.45em\lower1.2ex\hbox{\tiny
#2}}\;}

\def\cyclic{\mathop{\kern0.9ex{{+}
\kern-2.2ex\raise-.28ex\hbox{\Large\hbox
{$\circlearrowright$}}}}\limits}
\newcommand{\sech}{\mbox{sech}}

\newcommand{\fin}{\end{document}}

\def\rpartial{\mathrel{\partial\kern-.75em\raise1.75ex\hbox{$\rightarrow$}}}
\def\lpartial{\mathrel{\partial\kern-.75em\raise1.75ex\hbox{$\leftarrow$}}}

\def\beq{\begin{eqnarray}}
\def\eeq{\end{eqnarray}}

\def\m{\mu}

\def\cd{{\cal D}}
\def\ce{{\cal E}}
\def\cf{{\cal F}}

\def\cm{{\cal M}}

\def\co{{\cal O}}
\def\cp{{\cal P}}

\def\cs{{\cal S}}

\newcommand{\PP}{\mathcal P}

\newcommand\hsp[1] {\mbox{\hspace{#1 em}}}
\def\lefthook      {{\vrule height5pt width0.4pt depth0pt}}
\def\leftrighthookfill{$\mathsurround=0pt \mathord\lefthook
                   \hrulefill\mathord\righthook$}
\def\righthook     {{\vrule height5pt width0.4pt depth0pt}}

\newcommand\Cont[1]{\hsp{.4}\vtop{\ialign{##\crcr$\hfil\displaystyle
{\hsp{-.4}#1}\hfil\hsp{-.4}$\crcr\noalign{\kern-1.9pt\nointerlineskip\vskip2pt}
\leftrighthookfill\crcr}}\hsp{.4}}

\def \be{{\beta}}

\def \ra{{\, \rightarrow \,}}
\def \lra{{\longrightarrow}}

\def \p{{\partial}}

\def \C{{\mathbb C}}
\def \R{{\mathbb R}}

\newcommand{\re}[1]{(\ref{#1})}

\def\cd{{\cal D}}

\begin{document}

\title{The deformation quantizations of the hyperbolic plane}

 \author{P.~Bieliavsky \\{\it G\'eom\'etrie et Physique math\'ematique}\\ {\it
         Universit\'e Catholique de Louvain}\\ {\it
         2, Chemin du Cyclotron, B-1348 Louvain-la-Neuve, Belgium}\\ E-mail:
           {Pierre.Bieliavsky@uclouvain.be}
\\           
\\ S.~Detournay
         \\{\it INFN, Sezione di Milano}\\ {\it Via Celoria, 16, 20133 Milano, Italy}\\E-mail:
            {stephane.detournay@mi.infn.it}  
\\
\\  Ph.~Spindel \\{\it M\'ecanique et Gravitation}\\ {\it Universit\'e de Mons-Hainaut, 20
         Place du Parc}\\ {\it 7000 Mons, Belgium}\\E-mail:
		  {spindel@umh.ac.be}
		}

          \date{\today}
\maketitle

\begin{abstract}
We describe the space of (all) invariant deformation quantizations on the hyperbolic plane $\DD$ as 
solutions of the evolution of a second order hyperbolic differential operator. The construction is entirely explicit
and relies on non-commutative harmonic analytical techniques on symplectic symmetric spaces. The present work 
presents a unified method producing every quantization of $\DD$, and provides, in the 2-dimensional context, an exact solution to Weinstein's WKB quantization program within geometric terms. The construction reveals the existence of a metric of Lorentz signature canonically attached (or `dual') to the geometry of the hyperbolic plane through the quantization  process.
\end{abstract}
\section{Introduction}\label{Intro}
\subsection{Motivations}
The idea of the \emph{formal deformation quantization} program \cite{BFFLS}, initiated by Bayen, Flato, Fronsdal, Lichnerowicz and Sternheimer, is to generalize the Weyl product to an arbitrary symplectic (or Poisson) manifold. In this context, the framework of quantum mechanics is the same as classical mechanics, observables are the same, and quantization arises as a \emph{deformation} of the algebra of functions on the manifold, from a commutative to a non-commutative one.

\noindent A {\sl formal star product} (or {\sl deformation quantization}) of a symplectic manifold $(M,\omega)$
is an associative $\C[[\hbar]]$-bilinear product:
\begin{equation}
\star:C^\infty(M)[[\hbar]]\times C^\infty(M)[[\hbar]]\longrightarrow C^\infty(M)[[\hbar]]:(u,v)\mapsto u\star v
\end{equation}
such that, for $u,v\in C^\infty(M)$, the formal series
\begin{equation}\label{DefStar}
u\star v\;=\;\sum_{k=0}^\infty\,\hbar^k\,C_k(u,v)
\end{equation}
involves bi-differential operators $C_k:C^\infty(M)\times C^\infty(M)\to C^\infty(M)$
satisfying the following properties:
\begin{enumerate}
\item[(i)] $C_0(\,u\,,\,v\,)\;=\;u\,v$
\item[(ii)] $C_1(u,v)-C_1(v,u)\;=\;2i\{\,u\,,\,v\,\}$ where $\{\,,\,\}$ denotes the Poisson
bracket on $C^\infty(M)$ associated to $\omega$.
\end{enumerate}
Two such star products $\star^i\quad (i=1,2)$ on the same symplectic manifold $M$ are called {\sl equivalent} if there exists a formal series of the form 
\begin{equation}
T\;=\;I\,+\,\sum_{k=1}^\infty\hbar^k\,T_k
\end{equation}
where the $T_k$'s are differential operators on $C^\infty(M)$ such that for all $u,v\in C^\infty(M)$, one has
\begin{equation}
u\star^2v\;=\;T\,(\,T^{-1}u\star^1T^{-1}v\,)\;;
\end{equation}
the latter expression will be shortened by $\star^2=T(\star^1)$.

\noindent The first existence proofs were given in the 80's independently by Dewilde-Lecomte, Fedosov and 
Omori-Maeda-Yoshioka \cite{DW, Fe, OMY}. 
Equivalence classes of star products on a symplectic manifold are in 1-1 correspondence with the space
of formal series with coefficients in the second de Rham cohomology space of $M$ \cite{BCG, NT}.

\noindent The most important example of star product is the so-called Moyal star product $\star^0_\hbar $ on the plane $\R^2$
endowed with its standard symplectic structure $\Omega$:
\begin{eqnarray}\label{asympt}
 u \star^0_\hbar v &:=& u.v + i\frac \hbar 2\{ u,v \} +
 \underset{k=2}{\overset{\infty}\sum}\frac{(i\,\hbar/2)^k}{k!} \Omega^{i_1
 j_1}\cdots \Omega^{i_k j_k} \partial_{i_1\cdots i_k}u \partial_{j_1\cdots
 j_k}v \quad,\\
 \label{MoyalAsympt}
& = &\sum_{k=0}^{\infty} \, \frac{(i\,\hbar/2)^k}{k!}
\sum_{p=0}^{k}(-1)^{p} \,
\frac{k!}{p!(k-p)!}\, \partial_a^{k-p} \partial_\ell^p u \,
\partial_a^{p} \partial_\ell^{k-p} v  \quad ,
\end{eqnarray}
where \begin{equation}\label{PoissonBrackAN}
 \Omega_{ik}\Omega^{kj}=-\delta_i^k\qquad,\qquad \{ u,v \}=\Omega^{ij}\partial_iu\partial_jv: = \partial_a u \partial_\ell v - \partial_\ell u \partial_a
  v \quad.
  \end{equation}
\noindent In the context of formal deformation quantization, one does not worry about the convergence of the series (\ref{DefStar}). In particular, a formal star product does not, in general, underlie any operator algebraic or spectral theory. However, situations exist where star products occur as asymptotic expansions of operator calculi. The most known example is provided by the Weyl product which has the following integral representation:
\begin{equation}\label{IntMoy}
 (u \star^W v)(x) = (\pi\,\hbar)^{-2n} \int_{\R^{2n}\times \R^{2n}} \, u(y) v(z) \, e^{  i\frac{2}{\hbar} S^0(x,y,z)} dy dz,  
\end{equation}
where
\begin{equation}\label{Wphase}
  S^0(x,y,z)  =  \Omega (x,y) + \Omega (y,z) + \Omega (z,x), 
\end{equation}
$\Omega$ denoting the standard symplectic two-form on $\R^{2n}$. The product represented by \re{IntMoy} enjoys an important property: it is internal on the Schwartz space, the space of rapidly decreasing functions, on $\R^{2n}$\cite{Hansen}. Thus, the product of two (Schwartz) functions is again a (Schwartz) function ---rather than a formal power series as in the formal context. Such a situation will be referred to as {\sl non-formal deformation quantization}. Note that Moyal's star product (\ref{asympt}) can be defined as  a formal asymptotic expansion
of Weyl's product.

\noindent  In this paper, we will be interested in {\sl invariant} deformation quantizations. When a group $\LL$ acts on $M$ by symplectomorphisms, a star product is saaid to be $\LL$- {\sl invariant}
if for all $g\in\LL$, $u,v\in C^\infty(M)$:
\begin{equation}
g^\star(u\star v)\;=\;g^\star u\star g^\star v\;.
\end{equation}
When $\LL$ preserves a symplectic connection $\nabla$ on $M$:
\begin{equation}
\LL\subset\mbox{\rm Aff}(\nabla)\cap\mbox{\rm Symp}(\omega)\;,
\end{equation}
then, Fedosov's construction yields an $\LL$-invariant star product. The notion of equivalence
of $\LL$-invariant star products is the same as above except that each differential operator $T_k$
is required to commute with the action of $\LL$. The set of $\LL$-equivariant equivalence classes
of $\LL$-invariant star products is in this case parametrized by the space $H^2_{\mbox{\rm dR}}(M)^\LL[[\hbar]]$
of series with coefficients in the classes of $\LL$-invariant 2-forms on $M$ \cite{BBG}.

\vspace{2mm}

\noindent Although the study of deformation quantization admitting  a given symmetry is natural, one may 
mention important specific situations where the symmetric situation is relevant. 
Firstly, the observable algebra of a conformal field theory defines an algebra of vertex operators. Thus any finite dimensional limit of such a theory must contains some remanent of this structure. In particular if the limit procedure is equivariant the remanent will also be symmetric. For instance, in the seminal work of Seiberg and Witten [\cite{SeibWitt}] it is shown that , in the limit of a large B-field on a flat brane, the limit of the vertex operator algebra is an associative invariant product on the space of functions on the brane that has the symmetry of the initial brane and fields configuration. Accordingly it becomes natural to inquire about all the associative composition law of functions compatible with a given symmetry group (especially for non-compact symmetries). Actually, let us emphasize that our considerations go beyond string theory, but concern any Kac-Moody invariant theory.

\noindent Moreover, it is known since the seminal works \cite{ConnesDouglas,SeibWitt} that noncommutative geometry (and deformation theory) has intricate links with string theory. A manifestation of this statement stems from the fact that, in flat space-time, the worldvolume of a D-brane, on which open strings end, is deformed into a noncommutative manifold in the presence of a $B$-field. In particular, in the limit in which the massive open string modes decouple, the operator product expansion of open string tachyon vertex operators is governed by the Moyal-Weyl product\cite{SeibWitt, SchomDefQ}. This can be schematically written as
$ {\rm e}^{i P.X}(\tau)  {\rm e}^{i Q.X}(\tau') \sim ({\rm e}^{i P.X} \overset{M}{\star} {\rm e}^{i Q.X})(\tau')$, where $X$ represents the string coordinates on the brane, $P$ and $Q$ the momenta of the corresponding string states, $\tau$ and $\tau'$ parameterize the worldsheet boundary, and where the classical limit $f \overset{M}{\star} g \ra f.g$ is recovered for $B \ra 0$. It is not clear how this generalizes to an arbitrary curved string background supporting D-brane configurations. Nevertherless, some particular examples have been tackled in the litterature. The SU(2) WZW model for instance supports symmetric D-branes wrapping $S^2$ spheres in the $SU(2)$ group manifold. It has been shown that in an appropriate limit the worldvolume of these branes are deformed into fuzzy spheres \cite{AleksReckSchom2, AleksReckSchom1, AleksReckSchom3, Schomerus:2002dc}. The latter fuzzy spheres can be related to the Berezin/geometric quantization
of $S^2$ hence to star product theory (see \cite{PierreJan} and references therein). The appearance of noncommutative structures was also pointed out in relation with other backgrounds, like the Melvin Universe \cite{Hashimoto:2005hy,Hashimoto:2002nr,Halliday:2006qc,Behr:2003qc,Cai:2002sv,Cai:2006td,Cai:1999aw} or the Nappi-Witten plane wave \cite{Halliday:2006qc,Halliday:2005zt}, see e.g. \cite{Szabo:2006wx} for a recent review. Another model that has attracted much attention in recent years is the $\SL$ WZW model, describing string propagation on $AdS_3$ (and its euclidian counterpart, the $H_3^+$ model), see e.g. \cite{MO} and \cite{Teschner1}. This model has played, and still plays an important role in the understanding of string theory in non-compact and curved space-times. It appears as ubiquitous when dealing with black holes in string theory and is also particularly important in connection with the AdS/CFT correspondence. In this perspective, much effort has been devoted in studying the D-brane configurations this model can support (see e.g. \cite{Bachas:2000fr}). The most simple and symmetric D-branes in $AdS_3$ turn out to wrap $AdS_2$ and $H_2$ spaces in the $\SL$ group manifold. The question 
one could then ask is: what is the low energy effective dynamics of open string modes ending on such branes? By analogy with the flat case, it seems not unreasonable to expect a field theory defined on a noncommutative deformation of the D-branes' worldvolume. This deformation would have to enjoy some properties, namely to respect the symmetries of the model, just like the Moyal product and the fuzzy sphere construction do in their respective cases. 

\noindent In the present paper, we will be concerned with the specific situation of formal and non-formal deformation quantizations of  the hyperbolic plane. In this particular case, a third motivation relies on the relevance of the hyperbolic plane in the study of Riemann surfaces through the uniformization theorem (establishing the hyperbolic plane $\DD$  as the metric universal covering space of every constant curvature hyperbolic surface). Therefore, invariant deformations of the Poincar\'e disk could constitute a step towards a spectral theory of non-commutative Riemann surfaces.

\subsection{What is done in the present work}
We now summarize what is done in the present article as well as the method used. Given the geometric data
of an affine space (i.e. a manifold endowed with an affine connection), one defines the notion of {\sl contracted}
space as a pair constituted by the same manifold but endowed with a connection whose Riemann's curvature 
tensor appears as the initial one but where some components were `contracted' to zero. One then starts from the standard intuitive idea that every geometric theory (such as deformation quantization for instance) formulated at the 
level of the initial space turns into a simpler theory after the contraction process. In order to describe the theory at the initial level, one may try to describe the contracted theory first and then apply to the latter an operator that reverses the contraction process. This is exactly what is done here regarding deformation quantization of the hyperbolic plane. We observe that the hyperbolic plane admits a unique contraction into a generic co-adjoint orbit of the 
Poincar\'e group in dimension 1+1. The set of all Poincar\'e invariant deformation quantizations (formal and non-formal) of the latter contracted orbits was earlier bijectively  parametrized by an open subset of the algebra of pseudo-differential operators of the line  \cite{StarP, Bi07}. The correspondence as well as the composition
products were given there in a totally explicit manner. Both geometries (i.e. affine connections) on the hyperbolic
and contracted hyperbolic planes, although very different, share however a common symmetry realized by a
simply transitive action of the affine group of the real line: $\SS\simeq `ax+b'$.
From earlier results at the formal level, one knows that, up to a redefinition of the deformation parameter, two $\SS$-invariant star products are equivalent to each other under a convolution operator by a (formal) distribution $u$ on the Lie group $\SS$ \cite{BBG}. In the situation where one of them is $\mbox{SL}_2(\R)$-invariant and the other one is Poincar\'e-invariant, 
the distribution $u$ is shown to solve a second order canonical hyperbolic differential evolution equation.  Solving the latter evolution problem therefore reverses the contraction process, allowing to recover
the set of invariant star products (formal or not) on the hyperbolic plane from the set of contracted ones on the 
above mentioned Poincar\'e orbit. It is worth to point out that the Lorentz metric underlying the above Dalembertian,
that realizes {\sl per  se} the `` de-contraction" process, is canonically attached to the geometry of the (quantum) hyperbolic plane. Indeed, the latter metric does actually not depend on any choice made. To our knowledge,
the relevance of the above metric within hyperbolic geometry is new. The physical meaning of this canonical quantity has still to be clarified.

\noindent It turns out that the method of separation of variables applies to the above-mentioned evolution equation
yielding a space of solutions under a totally explicit form. Every solution $u$ can be realized as 
a superposition of specific modes $u_s$ given in terms of Bessel functions:
\begin{equation*}
u\;=\;\int_\R\,\tilde{u}(s)\,u_s\,{\rm d}s\;.
\end{equation*}
These modal
solutions $u_s$ provide non-formal invariant deformation quantizations of the hyperbolic plane. 
In particular, one of them ($u_0$) corresponds to a deformational version of Unterberger 's Bessel
calculus on the hyperbolic plane \cite{Unt,Unt2}.

\noindent From explicitness, one also deduces a geometric solution of Weinstein's WKB quantization program. 
The latter program proposes the study of  invariant star products on symplectic symmetric spaces
(see section \ref{Symsymspa}) expressed as an oscillatory integral, analogous to the oscillatory integral formulation (\ref{IntMoy}) of Weyl's composition. 
In \cite{WeinsteinTr}, A. Weinstein suggested a beautiful geometrical interpretation of the asymptotics of the phase $S=S(x,y,z)$ occurring in the oscillatory kernel in terms of the  area of a geodesic triangle admitting points
$x,y$ and $z$ as midpoints of its geodesic edges. In section (\ref{Unterberger}), we illustrate this by establishing an exact formula
for the kernel in terms of a specific geometrical quantity of three points hereafter denoted by $S_{\mbox{\rm can}}$,
in accord with Weinstein's asymptotics.

\section{Symplectic symmetric spaces}\label{Symsymspa}
\subsection{Definitions and elementary properties}\label{DefandProp}
Everything in this subsection is entirely standard and can be found e.g. in \cite{Bithese} and references therein.
A {\sl symplectic symmetric space} is a triple $(M,\omega,s)$ where $M$ is a connected smooth manifold
endowed with a non-degenerate two-form $\omega$ and where 
\begin{equation*}
s:M\times M\to M:(x,y)\mapsto s(x,y)=:s_xy
\end{equation*}
is a smooth map such that for all point $x$ in $M$, the partial map:
$s_x:M\to M$ is an involutive  diffeomorphism 
of $M$ (i.e. $s_x^2=\mbox{\rm id}_M$) which preserves the two-form $\omega$ (i.e. $s_x^\star\omega=\omega$)
and which admits $x$ as an isolated fixed point. On furthermore requires the following property:
\begin{equation*}
s_xs_ys_x\;=\;s_{s_xy}
\end{equation*}
to hold for any pair of points $(x,y)$ in $M\times M$.
In this situation, the space $M$ is endowed with a preferred affine connection. Indeed, 
for every triple of tangent vector fields $X,Y$ and $Z$ on $M$,
the following formula:
\begin{equation*}
\omega_x(\,\nabla_XY\,,\,Z\,)\;:=\;\frac12 X_x.\omega(\,Y\,+\,s_{x\star}Y\,,\,Z)
\end{equation*}
defines\footnote{See \cite{Bithese} and appendix \ref{AppenB} for an explicit computation in the case of generic coadjoint orbits of the Poincar\'e (1,1) group.} a torsion-free affine connection $\nabla$ on $M$ that enjoys the properties of being preserved by every 
{\sl symmetry} $s_x$ as well as being compatible with $\omega$ in the sense that:
\begin{equation*}
\nabla\omega=0\;.
\end{equation*}
This last fact implies in particular that $\omega$ is closed, turning it into a {\sl symplectic} form on $M$.

\vspace{2mm}

\noindent An important class of symplectic symmetric spaces is constituted by the non-compact {\sl Hermitean 
symmetric spaces}. Such a space is a coset space $M=G/K$ of a non-compact simple Lie group $G$ by
a maximal compact subgroup $K$ that admits a non-discrete center $Z(K)$. The first example
being the hyperbolic plane $\DD:=\mbox{SL}_2(\R)/\mbox{SO}(2)$.  The compactness of $K$ implies
in particular that a Hermitean symmetric space admits a $G$-invariant Riemannian metric for which
the connection $\nabla$ is the Levi-Civita connection. 

\noindent However, in general a symplectic symmetric space
needs not to be Riemannian, even not pseudo-Riemannian, in the sense that there is in general {\sl no} metric tensor $g$ on $M$ such that $\nabla g=0$. In this sense a symplectic symmetric space is a purely 
symplectic object. In the problem we are concerned with in the present work, such a ``non-metric"
symplectic symmetric space will play a central role.

\vspace{2mm}

\noindent Two symplectic symmetric spaces $(M_i,s^{(i)},\omega^{(i)})$ $(i=1,2)$ are said {\sl isomorphic}
is there exists a diffeomorphism $\varphi:M_1\to M_2$ that is symplectic i.e. such that $\varphi^\star\omega^{(2)}=\omega^{(1)}$ and that intertwines the symmetries: $\varphi s^{(1)}_x\varphi^{-1}=s^{(2)}_{\varphi(x)}$ for all $x$ in $M^{(1)}$. Given a symplectic symmetric space $(M,\omega,s)$,
its {\sl automorphism  group} $\mbox{\rm Aut}(M,\omega,s)$ therefore turns out to be the intersection of the diffeomorphism group
of affine transformation of $(M,\nabla)$ with the symplectic group of $(M,\omega)$:
\begin{equation}
\mbox{\rm Aut}(M,\omega,s)\;=\;\mbox{\rm Aff}(\nabla)\,\cap\,\mbox{\rm Symp}(\omega)\;.
\end{equation}
The latter group is therefore a (finite dimensional) Lie group of transformations of $M$. One 
then  shows that since it contains the symmetries $\{s_x\}_{x\in M}$ its action on $M$ is transitive,
turning $(M,\omega)$ into a {\sl homogeneous symplectic space}. In particular, every symplectic 
symmetric space is a coset space.

\vspace{2mm}

\noindent Up to isomorphism, the list of homogeneous spaces underlying simply connected 2-dimensional symplectic symmetric spaces is the following:
\begin{enumerate}
\item$\mbox{\rm the flat plane}: \R^2$
\item$\mbox{\rm the hyperbolic plane}: \DD\;:=\;SL_2(\R)/SO(2)$
\item$\mbox{\rm the universal covering space of the anti-de Sitter surface}: \widetilde{\mbox{\rm AdS}_2}:=\widetilde{SL_2(\R)/SO(1,1)}$
\item$\mbox{\rm the sphere}: S^2:=SO(3)/SO(2)$
\item$\mbox{the universal covering space of the Galileo coset}: \widetilde{SO(2)\times\R^2}/\R$
\item$\mbox{the Poincar\'e coset}: \MM:=SO(1,1)\times\R^2/\R$.
\end{enumerate}
Items 5 and 6 provide the first examples of non-metric symplectic symmetric spaces.

\noindent A old classical result independently due to Kirilov and Kostant \cite{Kos,Kir} asserts that every simply connected homogeneous symplectic
space is isomorphic to the universal covering space of a co-adjoint orbit of some Lie group. In our situation
of symplectic symmetric spaces this can be easily visualized. Indeed, fixing a base point $o\in M$ the conjugation
\begin{equation}
\tilde{\sigma}:\mbox{\rm Aut}(M,\omega,s)\to\mbox{\rm Aut}(M,\omega,s):g\mapsto\tilde{\sigma}(g):=s_ogs_o
\end{equation}
defines an involutive automorphism of the group $\mbox{\rm Aut}(M,\omega,s)$. Its differential at the unit
element therefore yields an involutive automorphism at the Lie algebra level:
\begin{equation}
\sigma:=\tilde{\sigma}_{\star e}:\mathfrak{aut}(M,\omega,s)\to\mathfrak{aut}(M,\omega,s)
\end{equation}
where $\mathfrak{aut}(M,\omega,s)$ denotes the Lie algebra of the automorphism group.
The latter Lie algebra therefore decomposes into a direct sum of ($\pm 1$)-eigenspaces for $\sigma$:
\begin{equation}
\mathfrak{aut}(M,\omega,s)\;=\;\kk\oplus\pp\;.
\end{equation}
Note that the differential at $e$ of the coset projection: $\pi:\mbox{\rm Aut}(M,\omega,s)\to M$
when restricted to the ($-1$)-eigensubspace $\pp$ provides a linear isomorphism with the tangent space at $o$:
\begin{equation}
\pi_{\star e}|_\pp:\pp\tilde{\longrightarrow}T_o(M)\;.
\end{equation}
The pull back of the symplectic structure at $o$ therefore defines a symplectic bilinear two-form on $\pp$:
\begin{equation}
\Omega\;:=\;(\pi_{\star e}|_\pp)^\star\,\omega_o\;.
\end{equation}
When extended by zero to the entire $\mathfrak{aut}(M,\omega,s)$ the element $\Omega$ is easily seen
to be a {\sl Chevalley 2-cocycle} in the sense that:
\begin{equation}
\Omega([X,Y],Z)+\Omega([Z,X],Y)+\Omega([Y,Z],X)=0\;.
\end{equation}
The 2-cocycle $\Omega$ needs in general not to be exact at the level of the automorphism algebra $\mathfrak{aut}(M,\omega,s)$. That is there does not always exist a linear form $\xi_o$
on $\mathfrak{aut}(M,\omega,s)$ such that 
\begin{equation}\label{XI}
\Omega(X,Y)\;=\;\xi_o[X,Y]\;.
\end{equation}
Note nevertheless that the only 2-dimensional non-exact case corresponds
to the flat plane $\R^2$. 

\noindent We now adopt the notation
\begin{equation}
G\;:=\mbox{\rm Aut}(M,\omega,s)\;;\quad\gg\;:=\;\mathfrak{aut}(M,\omega,s)\;.
\end{equation}
Denoting by $K$ the stabilizer subgroup of $G$ of the base point $o$, one gets the $G$-equivariant identification
\begin{equation}
G/K\longrightarrow M: gK\mapsto go\;,
\end{equation}
where the symmetry map reads
\begin{equation}
s_{gK}(g'K)\;=\;g\tilde{\sigma}(g^{-1}g')\,K\;.
\end{equation}
It turns out that the 2-cocycle $\Omega$ is exact if and only if the action of $G$ on $(M,\omega)$
is {\sl Hamiltonian} in the sense that, endowing $C^\infty(M)$ with the Lie algebra structure defined
by the Poisson bracket associated to the symplectic structure $\omega$, 
there exists a Lie algebra map
\begin{equation}
\lambda: \gg\longrightarrow C^\infty(M):X\mapsto\lambda_X
\end{equation}
that satisfies the following property:
\begin{equation}
\{\lambda_X,f\}(x)\;=\;\frac{d}{dt}|_{t=0}f(\exp(-t X)x)\;=:\;X^\star_x.f
\end{equation}
for all function $f\in C^\infty(M)$.
Note that the map $\lambda$ is then necessarily {\sl $G$-equivariant} in the sense that
\begin{equation}
\lambda_{\mbox{Ad}(g)X}(x)\;=\;\lambda_X(gx)
\end{equation}
for all $g\in G$, $X\in\gg$ and $x\in M$.

\noindent When Hamiltonian, one may choose:
\begin{equation}
\xi_o(X)\;:=\lambda_X(o)\;.
\end{equation}
In that case, denoting by $\co$ the co-adjoint orbit of the element $\xi_o\in\gg^\star$,
the {\sl moment mapping}:
\begin{equation}
J: M\longrightarrow\co\subset\gg^\star:x\mapsto[J(x):\gg\to\R:X\mapsto\lambda_X(x)]
\end{equation}
realizes a $G$-equivariant covering from $M$ onto the co-adjoint orbit $\co$. Note that the co-adjoint orbit
$\co$ is itself endowed with a canonical symplectic structure $\omega^\co$ defined at the level
of the fundamental vector fields for the co-adjoint action by
\begin{equation}
\omega^\co_\xi(X^\star,Y^\star)\;:=\;\xi[X,Y]\qquad(\xi\in\co)\;.
\end{equation}
With respect to the latter structure the moment map $J:M\to\co$ is symplectic. 

\vspace{2mm}

\noindent When non-exact, a passage to a central extension of  $\gg$ yields an entirely
similar situation. Indeed, the transvection group (generated by all the $s_x \circ s_y$ and forming a normal subgroup of Aut$(M)$ ) does not act in a strongly Hamiltonian manner on $M$.
However, one may consider the (non-split) central extension $0\to\R Z\to\tilde{\gg}:=\gg\oplus\R Z\to\gg\to0$ defined by $[X,Y]^{\sim}:=[X,Y]\oplus\Omega(X,Y)Z$, mimicking the passage from $\R^{2n}$ to the Heisenberg algebra in the flat situation. In this new set-up, $M$ may now be realized as a coadjoint orbit of the extended group and the above results remain valid under essentially the same form \cite{Bithese}.

\vspace{2mm}

\noindent To close this subsection, we observe that when the co-adjoint orbit $\co$ is simply connected
the moment mapping $J:M\to \co$ is then necessarily a global $G$-equivariant symplectomorphism. This turns out to be the case in items 1, 2, 4 and 6 in the above list of 2-dimensional spaces.

\subsection{Group type symplectic symmetric surfaces, curvature contractions}
A symplectic symmetric space $(M,\omega,s)$ is said to be {\sl of group type} if there exists in its
automorphism group $G$ a Lie subgroup $\SS$ that acts on $M$ in a simply transitive manner i.e.
in a way that for all $x$ in $M$ there is one and only one element $g$ in $\SS$ with $x=go$. In that
case, one has a $\SS$-equivariant diffeomorphism:
\begin{equation}
\SS\to M:g\mapsto go\;.
\end{equation}
The symplectic structure $\omega$ on $M$ then pulls back to the group manifold $\SS$ as a left-invariant
symplectic structure $\omega^\SS$. Also, the symmetry at $o$ corresponds to a symplectic involution
\begin{equation}
\Psi^M:\SS\to\SS
\end{equation}
that encodes at the level of $\SS$ the  whole structure of symmetric space of $(M,s)$:
the symmetry at a point $g\in\SS$ is given by:
\begin{equation}
s^M_g(g')\;:=\;g.\Psi^M(g^{-1}.g')\;.
\end{equation}
A quick look at the above list in the 2-dimensional case leads us to the observation that

\vspace{2mm}

\noindent{\sl up to isomorphism, there are two and only two non-flat symplectic symmetric affine geometries
that are of group type: the hyperbolic plane $\DD$ and the Poincar\'e orbit $\MM$}.

\vspace{2mm}

\noindent The corresponding automorphism subgroups are in fact both isomorphic to the 2-dimensional affine group $\SS$
that we realize as $\SS:=\{(a,\ell)\}$ with the group law:
\begin{equation}\label{ANmult}
(a,\ell).(a',\ell')\;:=\;(\,a\,+\,a'\,,\,e^{-2a'}\ell\,+\,\ell'\,)\;.
\end{equation}
The unit element is then $e=(0,0)$ and the inverse map is given by $(a,\ell)^{-1}=(-a,-e^{2a}\ell)$.

\noindent Within this setting (see appendix \ref{AppenB}), the affine structures are encoded by the maps :
\begin{equation}\label{PSIM}
\Psi^\MM(a,\ell)\;:=\;(\,-a\,,\,-\ell\,)\;;
\end{equation}
and 
\begin{equation}\label{PSID}
\Psi^\DD(a,\ell)\;:=\;\left(\,-a-\frac{1}{2}\log(1+\ell^2)\,,\,-\ell\,\right)\;.
\end{equation}

\noindent Regarding the symplectic structures, one observes that the constant 2-form ${\rm d}a\wedge{\rm d}\ell$
is invariant under the left-action. Therefore, for every $k\in\R^+_0$, the symplectic structure
\begin{equation}
\omega^{(k)}\;:=\;\sqrt{k}\,{\rm d}a\wedge{\rm d}\ell
\end{equation}
induces the following symplectic symmetric surfaces: $(\SS,s^\DD,\omega^{(k)})$ and $(\SS, s^\MM,\omega^{(k)})$.
The first one is isomorphic to the hyperbolic plane $\DD$ with curvature $\frac{-1}{k}$. In the second one,
the parameter $k$ is indifferent in the sense that for all $k$, one has the isomorphism: $(\SS, s^\MM,\omega^{(k)})\simeq(\SS, s^\MM,\omega^{(1)})$. 

\subsection{Admissible functions on symplectic symmetric surfaces}
In \cite{WeinsteinTr}, Alan Weinstein conjectures the relevance of a certain three-point function, here denoted
$S_W$, as the essential constituent of the phase of an oscillatory kernel defining an invariant star product
on the hyperbolic plane $\DD$ or more generally on any (reasonable) symplectic symmetric space $\co=G/K$. Essentially, when three points $x,y$ and $z$ in $\co$ are close enough to one another, Weinstein
function $S_W=S_W(x,y,z)$ is defined as the symplectic area of the geodesic (defined by the Loos connection) triangle
in $\co$ admitting points $x,y,z$ as mid-points of its geodesic edges.

\begin{figure}[h] 
   \centering
   \includegraphics[scale=1]{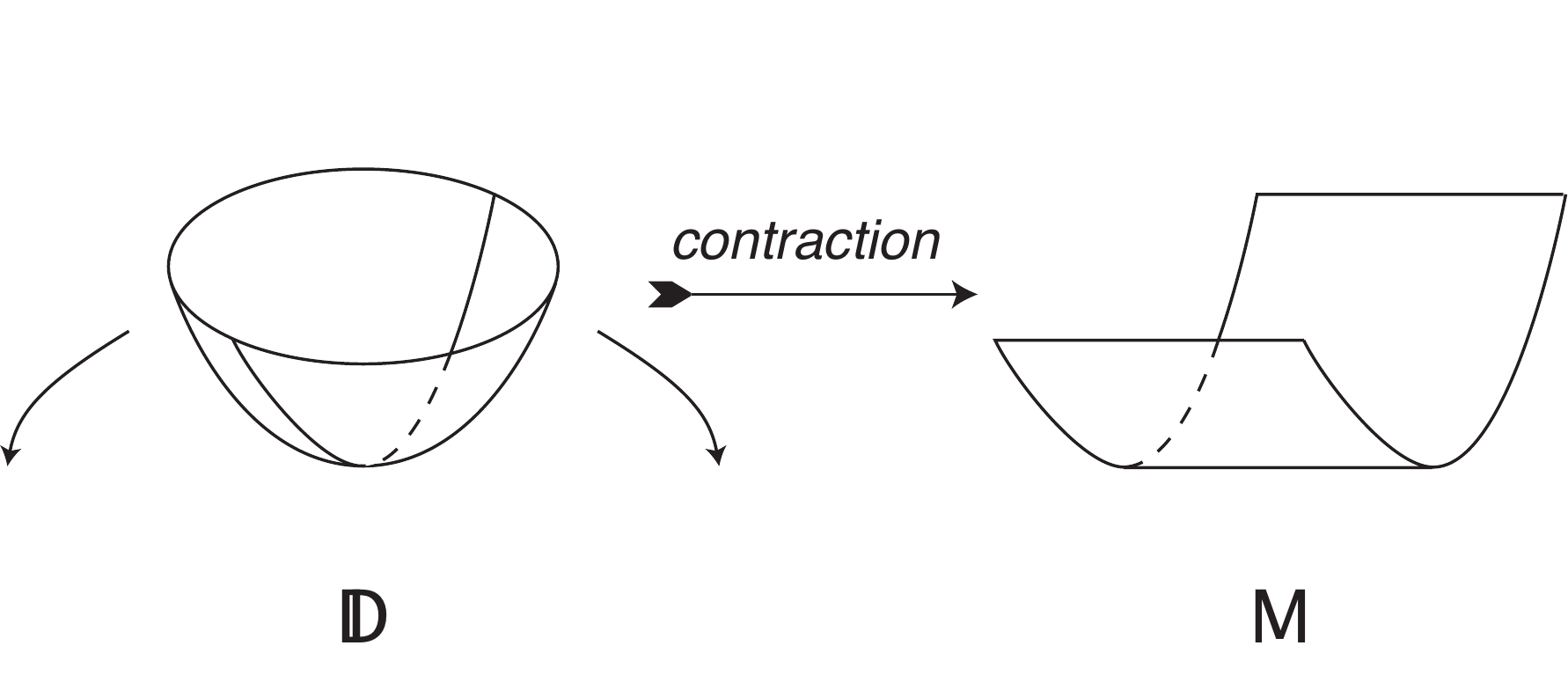} 
   \caption{Contraction of the Kaehler orbit into a non-metric hyperbolic cylinder.}
   \label{Figure 1}
\end{figure}

\noindent Additionally to its invariance under the symmetries, the three-point function $S_W$ (when defined)
has been shown in \cite{PierreStrict} to satisfy  the following so-called admissibility condition (\ref{ADM}). 

\noindent A three-point function
$S\in C^\infty(\co\times\co\times\co,\R)$ is called {\sl admissible} if it is invariant under the diagonal action of the symmetries on $\co\times\co\times\co$, is totally skewsymmetric, and if it satisfies the following property:
\begin{equation}\label{ADM}
S(x,y,z)=-S(x,s_{x}y,z)\;.
\end{equation}
As it will appear, the above condition is in fact the crucial one regarding star-products. 

\noindent In the case of a symplectic symmetric surface, every such (regular) admissible function turns out to 
coincide with an odd function of a {\sl canonical} admissible function. The latter function denoted hereafter $S_{\mbox{\rm can}}$ is defined in terms of the co-adjoint orbit realization of the symplectic symmetric surface
at hand. We now explain how to define it. 

\noindent Locally, every geodesic line
starting at the base point $o:=K$ and ending at $x$ can be realized as the (parametrized) orbit of $o$ by a one-parameter 
subgroup $\exp(tX)$ of $G$ with $X\in\pp$ ( see e.g. \cite{helgason} ). An invariant totally skewsymmetric smooth function $S\in C^\infty(\co^3)$ will be called {\sl regular admissible } if $S(o,x,y)=S(o,x,\exp(tX).y)$,
for all $t\in\R$ and $y\in\co$. Observe that regular admissibility implies admissibility as a consequence of the
following classical identity:
\begin{equation}
\exp(X)\;=\;s_{\mbox{\rm Exp}_o(\frac{1}{2}X)}\,\circ\, s_o\qquad(X\in\pp)
\end{equation}
where $\mbox{\rm Exp}_o$ denotes the exponential mapping at point $o$ with respect to the 
canonical connection.
The inclusion $\pp\subset\gg$ induces a linear projection  $\Pi:\gg^\star\to\pp^\star$. The dual space
$\pp^\star$ naturally carries a (constant) symplectic 2-form here denoted again by $\Omega$. Identifying 
points with vectors in $\pp^\star$, given two points $\xi$ and $\eta$, the quantity $\Omega(\xi,\eta)$
therefore represents the flat symplectic area of the (flat) Euclidean triangle in $\pp^\star$ admitting
points $0,\xi$ and $\eta$ as vertices. By transversality, the restriction  of the projection $\Pi$ to 
the co-adjoint orbit $\co\subset\gg^\star$ is locally (around $o$) a diffeomorphism.  Denoting by $\frac{x}{2}$ the unique point in a small neighbourhood of $o$ such that $s_{\frac{x}{2}}o=x$,
the following formula:
\begin{equation*}
S_{\mbox{\rm can}}(x,y,z)\;:=\;\Omega(\,\Pi(s_{\frac{x}{2}}(y))\,,\,\Pi(s_{\frac{x}{2}}(z))\,)
\end{equation*}
defines a (local) regular admissible function on $\co$. In the case where the surface $\co$ is of group type, the function $S_{\mbox{can}}$ is globally defined and smooth. 
Moreover every regular admissible function is an odd function of $S_{\mbox{can}}$. 

\begin{figure}[h] 
   \centering
   \includegraphics[scale=1]{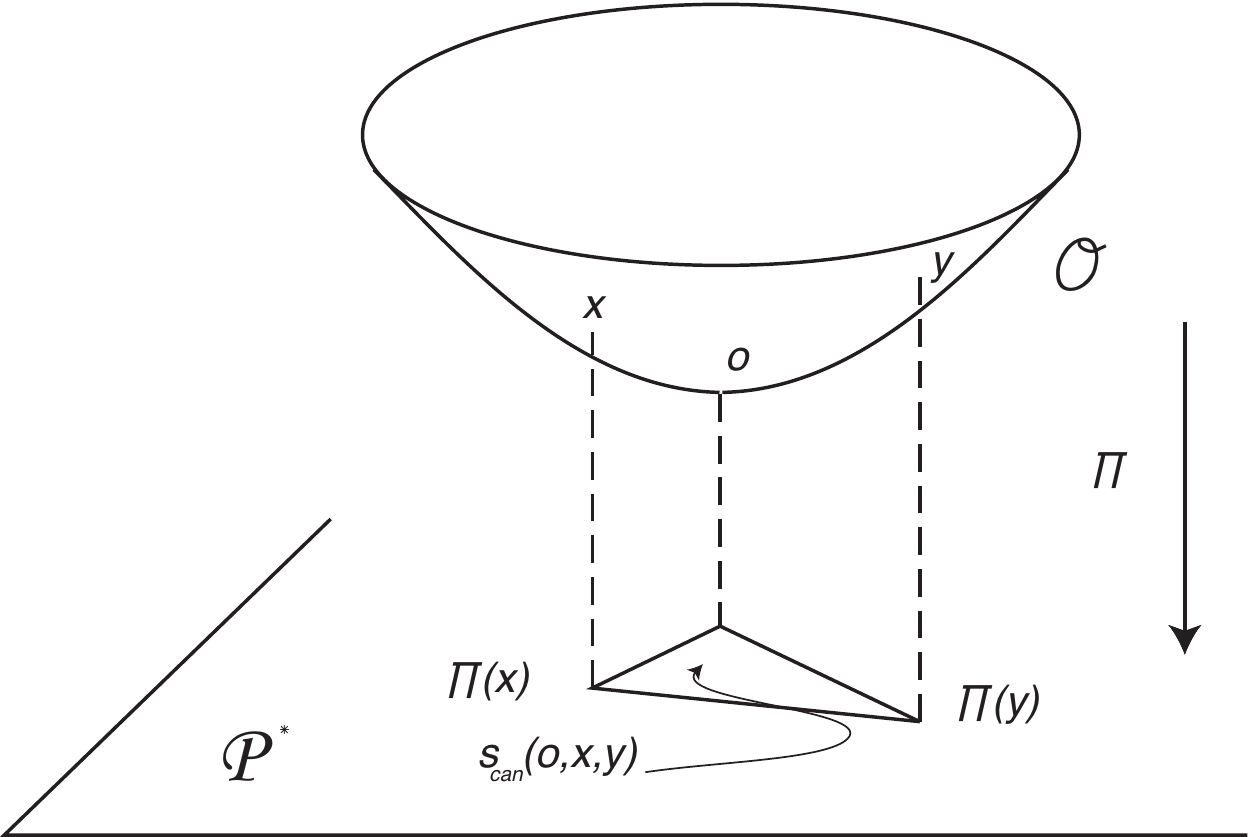} 
   \caption{The canonical three-point phase obtained from projecting the orbit.}
   \label{Figure 2}
\end{figure}
\noindent The proof essentially relies in the fact that the $\Pi$-projected $\exp(tX)$-orbits in $\co$ exactly
coincide with {\sl straight lines} in $\pp^\star$. Indeed, 
we first observe that the 
diffeomorphism $\Pi$ establishes a bijection between the 
$\exp(tX)$-orbits ($X\in\pp$) in $\co$ and the straight lines
in $\pp^\star$. Indeed, for $x\in\co$ and $X\in\pp$, one has $<\mbox{\rm Ad}^\star(\exp(tX))x-x,X>=0$, where 
$\mbox{\rm Ad}^\star$ denotes the co-adjoint action. Which means
that the $x$-translated  $\exp(tX)$-orbit of $x$ lies in the plane in $\gg^\star$ orthodual to $X\in\pp$.
This plane is generated by the kernel $\kk^\star$ of the projection $\Pi:\gg^\star\to\pp^\star$ and 
an element $X^\perp$ of $\pp^\star$ orthodual to $X$. In particular, it projects onto the line 
directed by $X^\perp$.
\noindent Now consider the two-point function $\kappa(x,y):=S(o,x,y)$ on $M$ induced by the data of an 
admissible function $S$ on $M$. This function corresponds to a two-point function $\kappa^0$ on $\pp^\star$
via the diffeomorphism $\Pi$. By admissibility and the above observation, one has
$\kappa^0(\xi,\eta)=\kappa^0(\xi,\eta+t\xi)$ for all $t\in\R$. Which is precisely the property of admissibility
for a two-point function with respect to the flat structure on $\pp^\star$. The rest then follows from Proposition 3.3 in \cite{PierreStrict}.

\vspace{3mm}

\noindent It is remarkable that in the case of Poincar\'e orbit $\MM$ the equation $s_xs_ys_zt=t$ admits a 
unique solution $t$ for all data of three points $x,y$ and $z$ in $\MM$. Moreover, the `double triangle' mapping\footnote{
Within A. Weinstein terminology.}:
\begin{equation*}
\Phi: \MM\times  \MM\times  \MM\to  \MM\times  \MM\times  \MM:(x,y,z)\mapsto(t,s_zt,s_ys_zt)
\end{equation*}
is a global diffeomorphism whose jacobian determinant equals
\begin{equation*}
\mbox{\rm Jac}_{\Phi}(x_{0},x_1,x_2)\;=\;16\,\cosh(2(a_0-a_1))\,\cosh(2(a_1-a_2))\,\cosh(2(a_2-a_0))\;.
\end{equation*}
\begin{figure}[h] 
   \centering
   \includegraphics[scale=1]{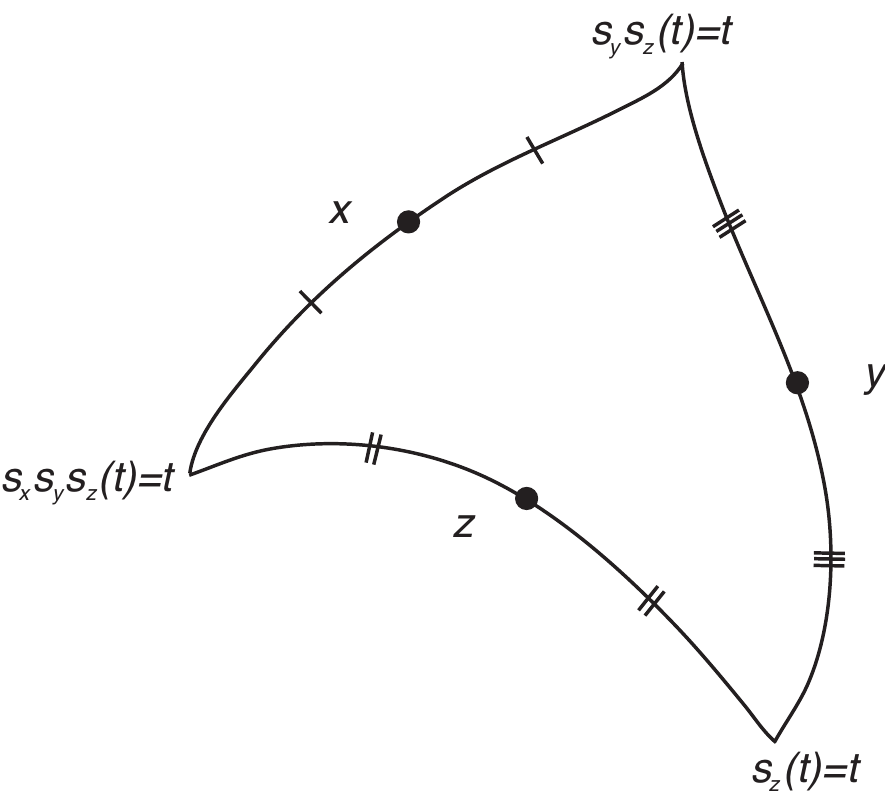} 
   \caption{The double geoedesic triangle.}
   \label{Figure 3}
\end{figure}

This is a straightforward computation based on the following formulas for the symmetries:
\begin{equation}\label{symPpl}
s^{\MM}_{(a,\ell)}(a',\ell')\,=\,(\,2a-a'\,,\,2\cosh(2(a-a'))\ell\,-\,\ell'\,)\;,
\end{equation}
as well as for the mid-point map:
\begin{equation}\label{midPpl}
m: \MM\times  \MM\to  \MM:(x,y)\mapsto m(x,y)=\left(\;\frac{1}{2}(a_x+a_y)\;,\;\frac{1}{2}(\ell_x+\ell_y)\sech(a_x-a_y)\;\right)\;,
\end{equation}
defined by the relation
\begin{eqnarray*}
s^{\MM}_{m(x,y)}x=y\;.
\end{eqnarray*}
One has 
\begin{equation*}
\Phi^{-1}(x,y,z)=(m(x,y),m(y,z),m(z,x))\;;
\end{equation*}
and a computation yields
\begin{equation*}
\mbox{\rm Jac}_{\Phi^{-1}}(x_0,x_1,x_2)=\frac{1}{16}\sech(a_0-a_1)\sech(a_1-a_2)\sech(a_2-a_0).
\end{equation*}
One then obtains the announced formula by using the relation: 
$\mbox{\rm Jac}_{\Phi}=(\Phi^\star\mbox{\rm Jac}_{\Phi^{-1}})^{-1}$.

\noindent At last, the canonical admissible three-point function, in this particular situation of $\MM$, exactly coincides
with Weinstein's function:
\begin{equation*}
S^{( \MM)}_{\mbox{\rm can}}\;=\;\mbox{ \rm Symplectic Area of } \Phi\;=\;S^{( \MM)}_W\;.
\end{equation*}
In coordinates, one has 
\begin{equation*}
S^{( \MM)}_{\mbox{\rm can}}(x_0,x_1,x_2)\;=\;\sinh(2(a_0-a_1))\,\ell_2+\,\sinh(2(a_2-a_0))\,\ell_1+\,\sinh(2(a_1-a_2))\,\ell_0\;.
\end{equation*}

\vspace{3mm}

\noindent In the hyperbolic plane case, however, the situation is not as nice. Indeed, a bit of reflection leads to the fact that Weinstein's
function is not well-defined for every triple of points $x,y$ and $z$ in $\DD$. In particular, since by construction $S^{(\DD)}_{\mbox{\rm can}}$ is smooth it must differ from $S^{(\DD)}_W$. Nevertheless, as proven above,
they should locally be odd functions of each other. Precisely, one has the relation:
\begin{equation*}
S^{(\DD)}_W\;=\;\pi-2\arccos(S^{(\DD)}_{\mbox{\rm can}})\;.
\end{equation*}
We set 
\begin{equation}\label{fW}
f_W(t)\;:=\; \pi-2\arccos(t)\;.
\end{equation}
In terms of the Kaehler potential $\zeta$ of the hyperbolic plane realized as the Poincar\'e unit disc, one has 
\begin{equation*}
S^{(\DD)}_{\mbox{can}}(0,z,w)\;=\;4\,\zeta(z)\,\zeta(w)\,\mbox{\rm Im}(z\overline{w})\,.
\end{equation*}
In coordinates, one has
\begin{equation}
S^{(\DD)}_{\mbox{can}}(0,x_0,x_1)=\ell_1 \sinh 2 a_0 -\ell_0\sinh 2 a_1 + \frac{\ell_0
\ell_1}{2}(\ell_0 e^{2 a_0} -\ell_1e^{2 a_1})\;.
\end{equation}
\noindent As we will see in the sequel, the 3-point kernel defining the associative  deformation product in the case of the hyperbolic plane turns out to be expressed as a special function (of one real variable) of the function $S_{\mbox{\rm can}}$ only (no approximation). While in the solvable contracted case, the latter coincides (up to a co-boundary) with the phase of the quantization kernel. This canonical function therefore appears as a unifying notion for  both contracted and un-contracted situations.

\section{Deformation Quantization}\label{QuaPoin}
\subsection{The space of star products on the Poincar\'e orbit $\MM:=SO(1,1)\times\R^2/\R$} It is proven in \cite{Bi07} that every Poincar\'e invariant star product on $\MM$ is realized as
a formal  asymptotic expansion in powers of $\hbar$ of an oscillatory integral expression in terms of the geometrical quantities defined above:
\begin{equation*}
u\,\star_{\hbar,{\cal P}} \,v\,(x_0)\;=\;\frac{1}{\hbar^2}\,\int_{\MM\times \MM}\,\sqrt{\mbox{\rm Jac}_{\Phi}(x_0,x_1,x_2)}\,e^{\frac{i}{\hbar}\,S^{(\MM)}_{\mbox{\rm can}}(x_0,x_1,x_2)}\,\frac{{\cal P}(a_0-a_1){\cal  P}(a_2-a_0)}{{\cal  P}(a_1-a_2)}\,u(x_1)\,v(x_2)\,{\rm d}x_1\,{\rm d}x_2\;;
\end{equation*}
where ${\rm d}x$ denotes the Liouville symplectic measure on $M$ and where $\cal P$ is an essentially arbitrary nowhere vanishing complex-valued one variable function possibly depending smoothly in the real deformation
parameter $\hbar$. The function $\cal P$
therefore being the only degree of freedom (see \cite{Bi07} for details).

\noindent We now briefly recall how the above result is obtained. The key point relies in the fact that the global Darboux coordinate system $(a,\ell)$ on $\SS$ enjoys a property of compatibility with 
the hyperbolic action of $G:=SL_2(\R)$ on $\DD=\SS$. 

\noindent In the case $\LL$ is a Lie group with Lie algebra $\mathfrak{l}$ that acts on a symplectic manifold $(M,\omega)$ in a Hamiltonian manner, a 
(not necessarily $\LL$-invariant) star product $\star$ on $(M,\omega)$  is called {\sl $\mathfrak{l}$-covariant}
if, denoting by
\begin{equation}
\lambda:\mathfrak{l}\longrightarrow C^\infty(M): X\mapsto \lambda_X
\end{equation}
the associated (dual) moment mapping, the following equalities hold:
\begin{equation}
\frac{i}{\hbar}\,[\,\lambda_X\,,\,\lambda_Y\,]_\star\;:=\;\frac{i}{\hbar}\,(\lambda_X\star\lambda_Y-\lambda_Y\star\lambda_X)\;=\;\{\,\lambda_X\,,\,\lambda_Y\,\}\;=\;\lambda_{[X,Y]}
\end{equation}
for all $X,Y\in\mathfrak{l}$. 

\noindent In this situation, one has a representation of $\mathfrak{l}$ on $C^\infty(M)[[\hbar]]$
by derivations of the star product $\star$:
\begin{eqnarray}
\rho_\hbar:\mathfrak{l}&\longrightarrow&\mathfrak{Der}(\star)\\
\rho_\hbar(X)\,u&:=&\frac{i}{\hbar}\,[\,\lambda_X\,,\,u\,]_\star\;.
\end{eqnarray}

\vspace{2mm}

\noindent It turns out that in coordinates $(a,\ell)$ the Moyal product (\ref{asympt}) is $\mathfrak{sl}_2(\R)$-covariant
with respect to the hyperbolic action of $G:=SL_2(\R)$ on the hyperbolic plane $\DD=\SS$ \cite{Bithese}. Precisely, presenting the Lie algebra $\gg:=\mathfrak{sl}_2(\R)$ as generated over $\R$ by $H, E$ and $F$ satisfying:
\begin{equation}
 [H,E] = 2E \quad,\quad [H,F] = -2F \quad,\quad [E,F] = H,\label{HEF}
 \end{equation}
the moment map associated with the action of $G$ on $\DD=G/K=\SS$ reads:
\begin{equation}\label{moments}
\lambda_H=\sqrt{k}\,\ell\,;\,\lambda_E=\frac{\sqrt{k}}{2}e^{-2a}\,;\,\lambda_F=-\frac{\sqrt{k}}{2}e^{2a}(1+\ell^2)\;.
\end{equation}
The associated fundamental vector fields are given by:
\begin{equation}\label{fundamental}
 H^\star = -\partial_a; \, E^\star = -e^{-2a}\partial_\ell;  \,
 F^\star= e^{2a}(\ell \partial_a - (k+\ell^2) \partial_\ell)\;.
 \end{equation}
At last, the representation of $\gg$ by derivations of $\star^0_\hbar$ admits the expression:
\begin{eqnarray}
\rho_{\hbar } (H) &=& -\partial_a \\
\rho_{\hbar }(E) &=& -\frac{e^{-2a}}{ \hbar } \sin ( \hbar \partial_\ell) \\
\rho_{\hbar }(F) &=&
 e^{2a}\left(\frac{\hbar  }4\sin ( \hbar \partial_\ell)\partial_a^2+
  \ell\cos ( \hbar \partial_\ell)\partial_a-(k+\ell^2)\frac{\sin ( \hbar \partial_\ell)}\hbar
\right)\;.\label{lambdaF} 
\end{eqnarray}
We now consider the partial Fourier transform in the $\ell$-variable
\begin{equation}
{\cal F}(\varphi)(a,\zeta) = \int_{-\infty}^\infty e^{-i \zeta \ell} \varphi(a, \ell) {\rm d}\ell \quad,
\end{equation}
and denote by $\tilde{\SS}:=\{(a,\zeta)\}$ the space where the Fourier transformed $\cf(g)$ is defined on.
\noindent Defining the following one-parameter family of diffeomorphisms of $\tilde{\SS}$:
\begin{equation}
\phi_\hbar  (a,\zeta) = \left(a,\frac{\sinh (\hbar\, \zeta)}{\hbar}\right) \quad ,
\end{equation}
and denoting by $\tilde{\cs}$ (resp. $\cs$) the space of Schwartz test functions $\cs(\tilde{\SS})$
(resp. $\cs({\SS})$) on $\tilde{\SS}=\{(a,\zeta)\}$ (resp. on $\cs(\SS)$ of $\SS=\{(a,\ell)\}$), one observes the following inclusions:
\begin{equation}
\phi_\hbar^\star\,\tilde{\cs}\,\subset\,\tilde{\cs}\;\mbox{\rm and }\,\tilde{\cs}\,\subset\,(\phi_\hbar^{-1})^\star\tilde{\cs}\,\subset\,\tilde{\cs}'\;.
\end{equation}
Therefore, every data of (reasonable) one parameter smooth family of invertible functions $\cp_\hbar=\cp_\hbar(b)$ yields an operator on the Schwartz space 
$\cs(\SS)$ of $\SS=\{(a,\ell)\}$:
\begin{equation}
T^{-1}:\cs(\SS)\longrightarrow\cs(\SS)
\end{equation}
defined as 
\begin{equation}\label{OpT}
T^{-1} \varphi(a_0,\ell_0)= \frac{1}{2\,\pi } \int \, e^{i\zeta \ell_0}\, \PP_\hbar(
\zeta)\, e^{\frac{-i}{\hbar}\ell \sinh(\hbar \zeta )} \, \varphi(a_0,\ell) \, {\rm d}\ell\,
{\rm d}\zeta\;.
\end{equation}
More generally, denoting by $\cm_f$ the pointwise multiplication operator by $f$, the  operator:
\begin{equation}
T:\cs(\SS)\longrightarrow\cs'(\SS)
\end{equation}
defined as
\begin{equation}
T\;:=\;\cf^{-1}\,\circ\,(\phi_\hbar^{-1})^\star\,\circ\cm_{\frac{1}{\cp}}\,\circ\cf
\end{equation}
is a left-inverse of $T^{-1}$.
Intertwining Moyal-Weyl's product by $T$ yields the above  product as $\star_{\hbar,{\cal P}} =T(\star^0_\hbar)$. Observe that the latter closes on the range space $\ce_\hbar:=T\cs$
yielding a (non-formal) one parameter family of associative  function algebras: $(\ce_\hbar,\star_{\hbar,{\cal P}} )$. Asymptotic expansions of these non-formal products produce
genuine Poincar\'e-invariant formal star products on $\MM$ \cite{BBM}.

\vspace{2mm}

\noindent For generic $\cal P$ the above oscillatory integral product defines, for all real {\sl value} of $\hbar$, an associative product law on some function space (as opposed to formal power series space) on $\MM$. The most remarkable case being probably the one where $\cal P$ is pure phase. In the latter case, the above
product formula extends (when $\hbar\neq0$) to the space $L^2(\MM)$ of square integrable functions as a Poincar\'e invariant Hilbert associative algebra. The star product there appears to be {\sl strongly closed}: for all
$u$ and $v$ in $L^2(\MM)$, $u\,\star_{\hbar,{\cal P}} \,v$ belongs to $L^1(\MM)$, and one has:
\begin{equation*}
\int_\MM u\,\star_{\hbar,{\cal P}} \,v\;=\;\int_\MM u\,v\;.
\end{equation*}

\vspace{3mm}

\noindent The contracted situation is therefore, to some extent, relatively well understood.

\subsection{$\mathfrak{sl}_2(\R)$- triplets of derivations and an unexpected Lorentzian structure}

\noindent Let us now consider {\sl any} Poincar\'e invariant formal star $\star$ product on the contracted plane $\MM$. Denote
by $\Der(\star)$ its algebra of derivations. Note that from formal equivalence with Moyal, every derivation is interior.
Consider in $\Der(\star)$ any element $D$ with the property that $D,H^\star$ and $E^\star$ form an $\mathfrak{sl}_2(\R)$-triplet of $\star$-derivations. The crucial point which the entirety of the present paper relies on resides now in the following totally unexpected fact:

\vspace{3mm}

\noindent {\sl Intertwining the derivation $D$ by the partial Fourier transform yields a second order hyperbolic differential operator $\Box\,:=\,\cf\,\circ\,D\,\circ\,\cf^{-1}$ whose principal symbol defines a Lorentzian metric which does not depend on any particular  choice made (of $\star$, $\mathfrak{sl}_2(\R)$-triplet derivation algebra and $D$)}.

\vspace{3mm}

\noindent The latter Lorentzian metric is therefore a new object that is canonically associated with the hyperbolic
plane, its canonical contraction and their quantization.

\vspace{3mm}

\noindent To prove the above assertion we first observe that the above covariance property implies that
the operator 
\begin{equation}\label{DERIV}
D_0\;:=\;T\circ\rho_\hbar(F)\circ T^{-1}
\end{equation}L
is a derivation of $\star:=T(\star^0)$. Now the particular choice of 
\begin{equation}
\cp(\zeta)\;\equiv1
\end{equation}
yields the following 
expression for $\Box:=\cf\,\circ\,D_0\,\circ\,\cf^{-1}$ that appears to be a second order differential operator:
\begin{eqnarray}\label{box1}
 \Box_{(a,\zeta)} = i e^{2a} \left[   \frac{\hbar^2}4\,\zeta\,
\partial_a^2 + \zeta\, (1+\hbar^2\, \zeta^2) \partial_\zeta^2 + (1+\hbar ^2
\, \zeta^2) \partial_a \partial_\zeta + \hbar^2 \,\zeta\, \partial_a  +  (2 + 3 \hbar^2 \, \zeta^2)
\partial_\zeta - \zeta\,(k-\hbar^2) \right] \;.\label{Box0}
\end{eqnarray}
Note the occurrence of the Lorentzian metric on $\tilde{\SS}$:
\begin{equation}
[g^{ij}]\;:=\;e^{2a}(1+\hbar ^2
\, \zeta^2) \,\left(
\begin{array}{cc}
 \frac{\hbar^2}4\,\frac{\zeta}{(1+\hbar ^2
\, \zeta^2) } & 1\\
1 & \zeta
\end{array}
\right)\;.
\end{equation}
Now, consider an arbitrary Poincar\'e invariant star product $\star$ on $\MM$. Note  that, denoting by $\ss$ the Lie algebra of $\SS$, the above Moyal-covariance property yields a $\ss$-quantum moment for $\star $ i.e. a linear map $\mathfrak{sl}_2(\R)\to C^\infty(\MM)[[\hbar]];X\mapsto\Lambda_X$ such that $\frac{1}{2\hbar}[\Lambda_X,\Lambda_Y]_\star=\Lambda_{[X,Y]}$ and $V^\star=\frac{1}{2\hbar}[\,\Lambda_V\,,\,.\,]$
for all $V\in\ss$. Fix $D_0$ in $\Der(\star)$ with the same property as $D$ and set $D=:D_0+D_1:=:D_0+[\lambda_1\,,\,.\,]$
with $\lambda_1\in C^\infty(\MM)[\hbar^{-1},\hbar]]$ (one knows from the equivalence with Moyal that $\Der(\star)$ is interior). The triplet condition yields the following conditions: $[E^\star,D_1]=0$ and $[H^\star,D_1]=-2D_1$.
The implies $\frac{1}{2\hbar}[\Lambda_E,\lambda_1]=E^\star.\lambda_1=c_E$ and $H^\star.\lambda_1=-2\lambda_1+c_H$ where $c_E$ and $c_H$ are formal constants. From the expressions (\ref{fundamental}),
one gets $\lambda_1=-c_Ee^{2a}\ell+const.$ i.e. $D=D_0+c_\hbar\,[e^{2a}\ell\,,\,.\,]_\star$ where $c_\hbar\in\C[\hbar^{-1},\hbar]]$. We now argue that the operator 
$\cf D_1\cf^{-1}:=\cf \circ c_\hbar\,[e^{2a}\ell\,,\,.\,]_\star\circ\cf^{-1}$ is differential and at most first order.
Indeed, starting with $D_0:=T^{-1}\rho_\hbar(F)T$ where $\star=T(\star^0)$, one observes by looking a the expression (\ref{moments}) of the classical moment that $D$ exactly corresponds to $\lambda_F$ affected by a 
translation in $\ell)$: $(a,\ell)\mapsto(a,\ell +c)$\footnote{Note that this symplectic transformation leaves Moyal' star product invariant.} (and a re-definition of $k$). From expression (\ref{lambdaF}), one 
deduces that $TD_1T^{-1}$ is a linear combination of $\cos(\hbar\partial_\ell)$,  
and $\ell\sin(\hbar\partial_\ell)$. The latter correspond under the $T$ equivalence to multiplication and 
vector fields operators. To complete the argument, we end by observing (see \cite{Bi07}) that one passes from one 
Poincar\'e invariant star product $\star'$ to $\star$ by an equivalence of the form $\cf^{-1}\circ\cm_{Q}\circ\cf$ where 
$\cm_Q$ denotes the multiplication operator by a one variable (formal) function $Q=Q(\xi)$ independent of $s$.
The latter has the effect of a simple gauge transformation affecting $\Box$ only by lower order terms.

\subsection{The de-contraction procedure: evolution of the Dalembert operator}
Let us now consider an invariant formal star product $\sharp$ on the hyperbolic plane $\DD$.
This star product is in particular $\SS$-invariant. One knows from \cite{BBG}, that the set 
of $\SS$-equivariant equivalence classes of $\SS$-invariant star products is in one-to-one correspondence with the set of  formal series with coefficients in the $\SS$-invariant second de Rham cohomolgy space. In the present two-dimensional situation, the latter space is simply
$\R[[\hbar]]$ since $H^2_{\mbox{\rm de Rham}}(\DD)^\SS$ is generated by the (non trivial) class
of the invariant symplectic structure (or area form). From \cite{BB}, one may therefore pass from one equivalence class of star products to another by re-defining the deformation parameter. In particular, up to a 
change of parameter, our star product $\sharp$ can be obtained by intertwining a given Poincar\'e
invariant star product $\star$ on the contracted plane $\MM$ through a formal equivalence $U$
that commutes with the left action of $\SS$ (identifying  $\DD$, $\MM$ and $\SS$). The equivalence $U$ between $\star$ and $\sharp=U(\star)$ must therefore be
a convolution operator by a (formal) distribution $u\in\cd'(\SS)$ on $\SS$ i.e. of the form:
\begin{equation}\label{conv}
(U f)(x) =  \int_{\SS\,}
u(y^{-1}x)\,f(y) \,{\rm d}^Ly\;,
\end{equation}
where ${\rm d}^Ly$ denotes a left-invariant Haar measure on $\SS$ (remark that it coincides with the
Liouville area form on $\SS=\DD=\MM$).

\noindent Now, on the one hand, the element $D:=D_F\,:=\,U^{-1}\circ F^\star\circ U$ is a derivation of $\star$ that generates
together with $E^\star$ and $H^\star$ a $\mathfrak{sl}_2(\R)$-triplet derivation algebra.
On the other hand the above subsection provides  us an explicit expression for  $D$, the identity
\begin{equation}\label{UMOINS1}
D\circ U^{-1}\;=\;U^{-1}\circ F^\star
\end{equation} 
may then be interpreted as an equation that must be satisfied by the intertwiner $U$ (or rather $U^{-1}$).
Denoting by $v$ the distribution on $\SS$ defining the kernel of $U^{-1}$, for any test function $f$, the latter  corresponds to
\begin{equation}
D_x\int v(y^{-1}x)f(y) \,{\rm d}^Ly=\int v(y^{-1}x)\,(F^\star_yf) \,{\rm d}^Ly\;,
\end{equation}
which yields, since $F^\star$ is a symplectic vector field:
\begin{equation}
\int D_x[v(y^{-1}x)]\,f(y) {\rm d}^Ly=-\int F^\star_y[v(y^{-1}x)]\,f(y) {\rm d}^Ly\;.
\end{equation}
Since the vector field $F^\star_y$ and the operator $D_x$ commute, the last equation leads to the following 
evolution equation for $D$:
\begin{equation}
D_x[v(y^{-1}x)]\;=\;-F^\star_y[v(y^{-1}x)]\;.
\end{equation}
In particular, we have shown that:

\vspace{2mm}

\noindent {\sl every $SL_2(\R)$-invariant star product $\sharp$ on the hyperbolic plane $\DD$ can be obtained by intertwining an arbitrary Poincar\'e invariant star product $\star$ on the contracted hyperbolic plane $\MM$ through a left invariant convolution operator $U^{-1}$ on $\SS$ whose associated kernel $v\in\cd'(\SS)$ is solution of the following problem:}
\begin{equation}\label{ES}
-\Box_{\tilde{z}} \,W(x,\tilde{z})\;=\;-F^\star_x\,W(x,\tilde{z})\;,
\end{equation}
with
\begin{equation}
W(\,x\,,\, (b,\zeta)\,)\;:=\;\int_{-\infty}^\infty e^{-i\zeta\ell} \,v(x^{-1}.(b,\ell))\,{\rm d}\ell\;;
\end{equation}
$(x\in\SS\,,\,\tilde{z}=(b,\zeta)\in\tilde{\SS})$.

\vspace{3mm}

\noindent We end this subsection by observing that the inverse map
\begin{equation}
\SS\longrightarrow\SS:x\mapsto x^{-1}
\end{equation}
induces a duality between the space of kernels $v$ of the above inverted intertwiners $U^{-1}$ and that 
of kernels $u$ defining the direct intertwiners $U$. To see this, we first observe that
given a Poincar\'e invariant star product $\star$ with associated trace form $\mbox{\rm Tr}_\star$, 
there exists a Poincar\'e equivariant equivalence:
\begin{equation}
\cd'(\SS)[[\hbar]]\longrightarrow\cd'(\SS)[[\hbar]]:\varphi\mapsto\underline{\varphi}
\end{equation}
such that every $\SS$-commuting intertwiner $U'$ may be expressed as
\begin{equation}
U'(\varphi)(x)\;=\;\mbox{\rm Tr}_\star\left(L_{x^{-1}}^\star\underline{u^\vee}\star\underline{\varphi}\right)\;;
\end{equation}
where $\varphi^\vee(x):=\varphi(x^{-1})$.
The latter being obvious for any strongly closed star product (with, in this case, $\underline{\varphi}=\varphi$), one gets it for every star product by use of an equivalence
with a strongly closed one. Right composition by Poincar\'e equivariant equivalences obviously preserves the space of $\SS$-commuting intertwiners. Hence, every $\SS$-commuting intertwiner $U$ may also be expressed as
\begin{equation}
U(\varphi)(x)\;=\;\mbox{\rm Tr}_\star\left(L_{x^{-1}}^\star\underline{u^\vee}\star\varphi\right)\;,
\end{equation}
for some $u\in\cd'(\SS)[[\hbar]]$.
\noindent Therefore, the equation $UD=F^\star U$ admits the expression:
\begin{equation}
\mbox{\rm Tr}_\star\left(L_{x^{-1}}^\star\underline{u^\vee}\star D\varphi\right)=\mbox{\rm Tr}_\star\left(F^\star_{(x)}(L_{x^{-1}}^\star\underline{u^\vee})\star \varphi\right)\;,
\end{equation}
The element $D$ being a derivation of $\star$, one has:
\begin{equation}
\mbox{\rm Tr}_\star\left(L_{x^{-1}}^\star\underline{u^\vee}\star D\varphi\right)=-
\mbox{\rm Tr}_\star\left(D(L_{x^{-1}}^\star\underline{u^\vee})\star \varphi\right)\;,
\end{equation}
which yields:
\begin{equation}
{D}(L_{x^{-1}}^\star\underline{u^\vee})=-F^\star_{(x)}(L_{x^{-1}}^\star\underline{u^\vee})\;.
\end{equation}
Now, from the particular form of Poincar\'e equivariant equivalence as power series
in the left-invariant $\tilde{E}$ with constant coefficient \cite{Bi07}:
\begin{equation}
\tilde{\tau}\;:=\;I\,+\,\sum_{k\geq1}c_k\,\hbar^k\,\tilde{E}^k\;,
\end{equation}
 one observes that
\begin{equation}
\underline{u^\vee}\;=\;\tilde{\tau}(u^\vee)\;=\;(\tau^\star(u))^\vee\;,
\end{equation}
with 
\begin{equation}
\tau^\star\;:=\;I\,+\,\sum_{k\geq1}c_k\,\hbar^k\,{E^\star}^k\;.
\end{equation}
\subsection{Finding solutions of the Dalembertian evolution by variable separation} \label{SectSL}
For convenience, we fix a particular choice of closed star product $\star$ on $\MM$ associated as above with
the function 
\begin{equation}
\cp(\zeta)\;=\;\cp^{\mbox{\tiny{\rm closed}}}(\zeta):=\sqrt{\cosh(\hbar \zeta)}\;.
\end{equation}
We denote the corresponding intertwiner by $T^{{\mbox{\tiny{\rm closed}}}}$. We then note that, from the above discussion,  the analogue of equation (\ref{ES}) for the direct operator $U$ in this case can be rewritten under the following form:
\begin{equation}\label{BoxBrutApp}
- \Box_{|\bar{{  z}}} W({  x},\bar{{  z}}) = F^\star_{{ (x)}}
W({ x},\bar{{  z}}) \quad ,
\end{equation}
with
\begin{equation}\label{DefWApp}
W({  x},\bar{{  z}}) = \left( (\phi_\hbar ^*)^{-1} \circ {\cal F}
\circ T^{{\mbox{\tiny{\rm closed}}}} \right)_{|{  z}} u({ z}^{-1}{  x}) \quad ,
\end{equation}
and where  the operator $\Box$ is given by (\ref{box1}). 
Observe that from  (\ref{DefWApp})
 the function $W({
x},\bar{{z}}) = W(a,n,b,r)$ takes the form 
\begin{equation}\label{FormeW}
W  (a,n,b,r) =   (1+ \hbar^2 \,r^2)^{1/4} \, q
\, e^{-i  r  n  q} \, F (q,r) \quad,
\end{equation}
with $q =  e^{2(a-b)}$ and
\begin{equation}\label{uF}
F (q,r) = \int\, e^{i\,  q\,r\, y} \, u
(\frac{1}{2} \ln  q,y) \, dy \quad,
\end{equation}
and where $ (a,n)$ are the coordinates of $x$, $(b,\ell)$ those of $z$, and, 
$(b,r)$ those of $\bar z$, $r$ being the variable conjugated to $\ell$ through the
Fourier transformation  ${\cal F}$ which only affect the second
$z$-coordinate.

\noindent The right-hand side of eq. (\ref{BoxBrutApp}) is simpler  :
\begin{eqnarray}\label{BoxBox}
F^\star_{ x}W(x,\bar z)= e^{2\,a}q\left(2\,q\,n\,\partial_q-(k'+n^2)\partial_n\right) W(q,n,r)\quad 
\end{eqnarray}
$k'$ being for the operator $F^\star_{x}$ the corresponding of $k$ for the  operator $F$ in $\rho_\hbar ^1(F)_{|{  z}}$  (see eqs (\ref{fundamental})), so finally eq. (\ref{BoxBrutApp}) simplifies to the following equation for $F  (q,
r)$ :
\begin{equation}\label{BoxFqr}
\left(-r\,  \hbar^2 q^2\,(1+\hbar^2  r^2  ) \partial_q^2 - r\,(1+\hbar^2 r^2)^2 \partial_r^2  + 2\,q\,(1+\hbar^2 r^2)^2 \partial_r
\partial_q   - 2\,\hbar^2\,r^2\,(1+\hbar^2 r^2)  \partial_r +
V\right) F (q,r) = 0 \ ,
\end{equation}
with 
\begin{equation}
V\;=\;V(q,r)\;:=\;r\,(-\frac{\hbar^2}4 (2+\hbar^2 r^2 ) + (1+\hbar^2 r^2) (k - k' q^2))\;.
\end{equation}

\noindent The problem  amounts thus to solve the corresponding equation for $F_\hbar (q,
r)$, from which we could find $u_\hbar$ by
inverting (\ref{uF}). We set 
\begin{equation}
G(q,r)  = q^{1/2} (1+ r^2\hbar^2)^{1/4}\,F(q,r)\;,
\end{equation}
\noindent and use the variables:
\begin{equation}\label{UVqr}
U = q^{\frac{1}{2}}\sinh(\frac{\hbar}{2}\zeta) \quad, \quad V =
q^{\frac{1}{2}}\cosh(\frac{\hbar}{2}\zeta)  \quad,
\end{equation}
with
\begin{equation}\label{rzeta}
r\;=:\;\frac{1}{\hbar}\sinh(\hbar\zeta)\;.
\end{equation}
\noindent The equation then becomes
\begin{equation}\label{wave}
\left[\partial_U \partial_V + \tilde{Q}(U,V)\right] G(U,V) = 0
\quad,
\end{equation}
with
\begin{equation}
\tilde{Q}(U,V) = -\frac{4}{\hbar^2} \frac{UV}{(U^2 - V^2)^2}
(\frac{\hbar^2}{4} - k + k'(U^2 - V^2)^2) \quad.
\end{equation}
Note that here : $U^2 - V^2 = -q$, hence $U^2 - V^2<0$. Setting  $U^2=\frac{1}{2}(t-x)$,
$V^2=\frac{1}{2}(t+x)$ and $G(q,r)=H(t,x)$, one then gets
\begin{equation}
\left\{x^2(\partial^2_t-\partial^2_x)-\frac
1{\hbar^2}[\frac{\hbar^2}{4} -k+k'x^2]\right\}H(t,x)=0\;.
\end{equation}
Applying a method of separation of variables to the latter yields solutions  as  superpositions of 
the following modes:
\begin{equation}
F_{\hbar,s} (q,r) =  h_s(q) (1+\hbar^2
\, r^2)^{-1/4} e^{i s q\sqrt{1+\hbar^2\,r^2} }\quad,
\end{equation}
with
\begin{equation}\label{hs}
h_s(x)= A\, J_{\frac
{\sqrt{k}}{\hbar}}[\sqrt{s^2+ {k'} /{\hbar^2}}\,\,
x] + B\, Y_{\frac{\sqrt{k}}{\hbar}}[\sqrt{s^2+  {k'}/ {\hbar^2}}\,\, x]\quad,
\end{equation}
where $J_\m$ and $Y_\mu$ denote  Bessel functions of the first and second kinds (see ref. \cite{Wa}).
From (\ref{uF})
\begin{equation}\label{Ftou}
u(a,l) = \frac{e^{2a}}{2 \pi}\int\, e^{-i \,r\, l\, e^{2a}} F(
e^{2a},r) \, dr \quad,
\end{equation}
thus (up to constant factors)
\begin{equation}\label{uGen}
u_{\hbar,s}(a,l) =  e^{2a} \, h_s(  e^{2a})\, \int \, {\cal
H}_s(a,b) \, e^{i  b  l e^{2a}} \, db \quad,
\end{equation}
with
\begin{equation}\label{uGenbis}
{\cal H}_s(a,b) = (1+ \hbar^2r^2)^{-1/4} e^{  i s
\sqrt{1+\hbar^2r^2} e^{2a}} \quad.
\end{equation}
We may use a similar technique to determine the kernel of the inverse operator $U^{-1}$. By writing
\begin{equation}
 [U^{-1}f]({x})=\int \, v({z}^{-1}{x}) \, f({z}) d{z} \quad,
 \end{equation}
one finds:
\begin{equation}\label{BoxBrutAppInv}
 \Box_{\bar{{x}}} M(\bar{{x}},{z}) = - F^\star_{{z}} M(\bar{{x}},{z}) \quad
 \end{equation}
with
\begin{equation}\label{M}
M(\bar{{  x}},{z}) = \left( (\phi_\hbar^*)^{-1} \circ {\cal F}
\circ T^{{\cal P}} \right)_{|{ x}} v^{{\cal P}}({z}^{-1}{x}) \quad .
\end{equation}
Solutions to (\ref{BoxBrutAppInv}) can be deduced by using a slight generalization of what we have done here above. We finally get,
\begin{equation}\label{vGen2}
v_{\hbar,s}(a,l) = e^{-2a} \, h_s(  e^{-2a})\, \int
\, (1+ \hbar^2r^2)^{-1/4} e^{  i s \sqrt{1+\hbar^2r^2} e^{-2a}} \, e^{-i b l} \, db \quad.
\end{equation}
\subsection{Normalizations and asymptotic expansions}\label{Norasympt}
\noindent Observe that the   change of variables (\ref{UVqr}) becomes singular when $\hbar$ vanishes, while the  equation (\ref{BoxFqr}) degenenrates into
\begin{equation}\label{BoxFqrdeg}
(-r\,\partial_r^2+2\,q\,\partial_r\partial_q+r\,(k-k'q^2))F(q,r)=0\qquad .
\end{equation}
The general solution of this equation can be expressed as a superposition of modes
\begin{equation}
F_{0,\sigma}=C\,\frac 1{\sqrt q}\,e^{i\frac{\sigma\,q\,r^2}2}\,e^{-\frac i{2\,\sigma}\left(\frac k q +k'q\right)}\qquad.
\end{equation}
Let us notice that these modes can be obtained from special combinations of modes (\ref{hs}) and a rescaling of the wave-number $s$ into $s=\sigma/\hbar^2$.
This rescaling is dictated by the necessity of maintaining a $r$ dependence in the limit $\hbar \rightarrow 0$ of the modes (\ref{hs}):
\begin{equation}
s\,\sqrt{1+\hbar^2r^2}\approx s + s\,\hbar^2r^2=\frac \sigma{\hbar^2}+\sigma \,r^2
\end{equation}
while the oscillating factor can be controlled by considering the asymptotic behavior of the  Hankel functions (\cite{Wa}, sec. {\bf 8.41})
\begin{equation}
H^{(2)}_{\frac{\sqrt k}{\hbar}}(\sqrt{\frac{\sigma^2}{\hbar^4}+\frac{k'}{\hbar^2}}q)\approx \hbar \sqrt{\frac 2{\pi\,\sigma\,q}}\,e^{-i\frac{\sigma\,q}{\hbar^2}}
e^{i\frac{\sqrt k\,\pi}{2\,\hbar}}\,e^{-\frac i{2\,\sigma}\left(\frac k q +k'q\right)}\qquad .
\end{equation}
Thus by choosing in the coefficients $A$ and $B$ in the linear combination (\ref{hs}) as :
\begin{equation}
A=-i\,B=C\,\frac 1 \hbar\,\sqrt{\frac{\pi\,\sigma}2}\,e^{-i\frac {\sqrt{k}\,\pi}{2\,\hbar}}
\end{equation}
we obtain   modes whose limit for $\hbar =0$ is well defined. This explain the results of our computations below.
 Note that, given an operator $U$, it is not straightforward to extract the inverse operator $U^{-1}$ since one has to determine how to superpose these different modes. In section \ref{Unterberger} however, we will discuss a particular example where the superposition is known.

\noindent The operators $U$ we consider hereafter will correspond  to \re{uGen}, \re{hs} with $B=0$. This special choice is motivated by the bad behavior of the $Y$-Bessel function near the origin. We have :
\begin{eqnarray}\label{USF}
 (U_s f)(a,l) =  C_\hbar (s) \int_{\R^2\times\R_+} \, dr \, dn\, dx\, J_{\frac{\sqrt{k}}{\hbar}}(\sqrt{s^2 + k'/\hbar^2}\, \,x) (1+\hbar^2 \,r^2)^{-\frac{1}{4}} \nonumber\\
  e^{i \,s\, x \sqrt{1+\hbar^2\, r^2}} \, e^{i \,r\, l x} \, e^{-i\, r\, n} \, f(a-\frac{1}{2}\ln x,n)\qquad.
\end{eqnarray}
A necessary condition for $U_s$ to define a star product is that $(U_s 1) = 1$. This leads to the condition\footnote{A useful relation in this derivation (see ref.\cite{Wa}, sec. 13.2, eq.[8]) is :
$
\int_0^\infty e^{(i\kappa-\epsilon)\zeta}J_\lambda (\beta\,\zeta)\,d\zeta=\frac{\beta^{-\lambda }[\sqrt{\beta^2-\kappa^2}+i\kappa]^{\lambda }}{\sqrt{\beta^2-\kappa^2}}
$}
\begin{equation}\label{Cnorm}
  C_\hbar(s) = \frac{ \sqrt{k'}}{2\,\pi\,\hbar}\left[ \frac{\sqrt{k'} - i \,\hbar\,s[\hbar]}{\sqrt{k'} + i \,\  \hbar\,s[\hbar]} \right]^{\frac{\sqrt{k'}}{2\,\hbar}}.
\end{equation}
Let us notice that we allow a $\hbar$ dependence in  $s$; we don't assume it a priori to be constant but  leave it as a free (regular) function of the deformation parameter.

\noindent On the other hand, the requirement 
\begin{equation}
   U_s f \underset{\hbar \ra 0}{\lra} f
\end{equation}
to get the right classical limit further imposes
\begin{equation}\label{kmu}
 k = k' \qquad .
\end{equation}
Let us now turn to the asymptotic expansion of the operator  \re{USF}. To this end we introduce the following Fourier-like transforms:
\begin{eqnarray}
 \tilde{f}_- (k,p) &=& \frac{1}{2\pi^2} \int f(a,l) e^{-i k e^{-2a}} e^{-ipl} e^{-2a} da\, dl \label{Fouriermoins}\\
  f(a,l) &=& \phantom{\frac{1}{2\pi^2}}\int \tilde{f}_- (k,p) e^{i k e^{-2a}} e^{i p l} dk dp\qquad ,
\end{eqnarray}
from which one obtains
\begin{eqnarray}\label{UsDev}
 (U_s f) (a,l) =\sqrt{k}\int \frac{(1+\hbar^2\, p^2)^{-1/4}}{(k + \hbar^2\,(s^2(\hbar) - \kappa^2))^{1/2}} \left[\frac{\sqrt{k + \hbar^2 \,(s^2(\hbar) - \kappa^2)} + i \,\hbar\,\kappa }{\sqrt{k} +  i \, \hbar\, s(\hbar)} \right]^{\frac{\sqrt{k}}{ \hbar}}  \tilde{f}_- (r,p)\,dp\,dr
\end{eqnarray}
where 
\begin{eqnarray}
 \kappa = r\, e^{-2a} + p\, l  + s(\hbar)\, \sqrt{1+\hbar^2\,r^2}.
\end{eqnarray}
Setting $X = r\, e^{-2a} + p\, l$, the integrand can be Taylor expanded around $\hbar = 0$ as 
\begin{equation}\label{TaylorU}
 \sqrt{k}  \frac{(1+\hbar^2\, p^2)^{-1/4}}{(k + \hbar^2\,(s^2(\hbar) - \kappa^2))^{1/2}} \left[\frac{\sqrt{k + \hbar^2 \,(s^2(\hbar) - \kappa^2)} + i \,\hbar\,\kappa }{\sqrt{k} +  i \,\hbar\, s(\hbar)} \right]^{\frac{\sqrt{k}}{ \hbar}}= {\rm e}^{i X} P(X,p^2)
\end{equation}
where \re{TaylorU} defines the $\hbar$ power series  $P(X,p^2)$,  which, at first (non-trivial) order\footnote{If we assume $s$ to be constant, instead of a function of $\hbar$, this expansion involves only even power of $\hbar$.}, reads as :
\begin{equation}\label{P}
 P(X,p^2) = 1 + \hbar^2 \left(-p^2\frac{ (1-2\,i\, s)}4 + \frac{ s}{k} X +  \frac{(1+ i\, s)}{2\,k} X^2 +  \frac{i}{6\, k}X^3\right) + O(\hbar^4) 
\end{equation}
The important point in the previous expression is the structure of the expansion, and in particular the occurrence of the factor $ {\rm e}^{i X}$. Using this result, the expansion of the operator $U$ follows:
\begin{eqnarray}
 (U_s f)(a,l) &\approx& \int P(X,p^2) e^{i X} \tilde{f}_- (r,p)\,dp\,dr \nonumber \\
              &=& \int   P[-i \p_\sigma,-\p^2_\alpha]\, e^{i \sigma X} e^{i\alpha p}\left . \tilde{f} (k,p)_- \right \vert_{\sigma=1, \alpha=0} \,dp\,dr  \nonumber \\
              &=& P[-i \p_\sigma,-\p^2_\alpha]\left. f(a-\frac{1}{2} \ln \sigma, \sigma l + \alpha)\right\vert_{\sigma=1, \alpha=0}\quad .
\end{eqnarray}
It therefore appears appears that the product so-defined deforms the pointwise product on the hyperbolic plane in the direction of the $\SL$-invariant Poisson bracket.

\noindent A similar computation can be performed starting from the inverse operator. One has (with $x= e^{2(b-a)}$)
\begin{eqnarray}
 (V_s f)(a,l) 
  =\frac{D_\hbar (s)}{2}\int_{\R^2\times\R_+} \, dr \, dn\, dx\, J_{\frac{\sqrt{k}}{\hbar}}(\sqrt{s^2 + k'/\hbar^2} \,x) (1+\hbar^2 \,r^2)^{-\frac{1}{4}} \nonumber\\
  e^{i \,s\, x \sqrt{1+\hbar^2\, r^2}} \, e^{ i r ({n\,x} -l)} \,   f(a+\frac{1}{2}\ln x,n)\qquad.\label{Vsf}
\end{eqnarray}
The requirement $V_s 1 = 1$, imposes
\begin{equation}\label{Dnorm}
 D_\hbar (s)=\frac{\sqrt{k}}{ \pi\,\hbar}\left(\frac{\sqrt{k'}-i\,\hbar\,s(\hbar)}{\sqrt{k'}+i\,\hbar\,s(\hbar)}\right)^\frac {\sqrt{k}} {2\,\hbar} 
 \end{equation}
while the right classical limit implies \footnote{A useful relation in that case is (see ref.\cite{Wa}, sec. 13.2, eq.(7)) : $
\int_0^\infty \frac{e^{(i\kappa-\epsilon)\zeta}} \zeta J_\lambda (\beta\,\zeta)\,d\zeta=\frac{[\sqrt{\beta^2-\kappa^2}+i\kappa]^{\lambda }}{\lambda \,\beta^{\lambda }}$}
\begin{equation}\label{kmu2}
  k =  k',
\end{equation}
in accordance with eq. \re{kmu}. 

\noindent Using again the expression of the Fourier transform of the Bessel function but   the Fourier transform
\begin{eqnarray}
 \tilde{f}_+ (k,p) &=& \frac{1}{2\pi^2} \int f(a,l) e^{-i k e^{2a}} e^{-ipl} e^{2a} da\, dl \label{Fourierplus}\\
  f(a,l) &=& \phantom{\frac{1}{2\pi^2}}\int \tilde{f}_+ (k,p) e^{i k e^{+2a}} e^{i p l} dk dp.
\end{eqnarray}
 one is led to
\begin{eqnarray}\label{VsDev}
 (V_s f)(a,l)  =\frac{\sqrt{k}}{2\,\pi}\int \frac{(1+\hbar^2\, p^2)^{-1/4}\,e^{i p l}\, e^{i \,q\, n}}{(k +   \hbar^2\,(s^2(\hbar) - \rho^2))^{1/2}} \left[\frac{\sqrt{k +  \hbar^2 \,(s^2(\hbar) - \rho^2)} + i  \,\hbar\,\rho }{\sqrt{k} +  i \, \hbar\, s(\hbar)} \right]^{\frac{\sqrt{k}}{ \hbar}}  \tilde{f}_+ (r,q)\,dp\,dr\,dn\,dq
\end{eqnarray}
where
\begin{equation}
 \rho =  {r\, e^{2a} - p\, n}  + s(\hbar)\, \sqrt{1+\hbar^2 \,p^2}\qquad.
\end{equation}
Setting $Y = r\, e^{2a} - p\, n$, we may write
\begin{equation}
\sqrt{k} \frac{(1+\hbar^2\, p^2)^{-1/4} }{(k +  \hbar^2\,(s^2(\hbar) - \rho^2))^{1/2}} \left[\frac{\sqrt{k +  \hbar^2 \,(s^2(\hbar) - \rho^2)} + i \, \hbar\,\rho }{\sqrt{k} +  i \,\hbar\, s(\hbar)} \right]^{\frac{\sqrt{k}}{ \hbar}} = e^{i Y} P(Y,p^2),
\end{equation}
where $P$ has the same expression as in \re{TaylorU} (Let us remark that this holds only  because we use the strongly closed $\SS$-invariant star product).
One successively gets
\begin{eqnarray}
 (V_s f)(a,l)  &=& \frac{1}{2\pi} \int dp \, dr\, dn \,dq \, e^{i \,p\, l} \,e^{i\,q\,n} P(Y,p^2) e^{iY} \tilde{f} _+(r,q) \nonumber \\
               &=&   \frac{1}{2\pi} \int dp \, dr\, dn \,dq \,  P[-i \p_\sigma,-\p^2_l] e^{i \sigma Y} \, e^{i \,p\, l} \,e^{i\,q\,n}\left . \tilde{f}_+ (k,q)\right\vert_{\sigma=1} \nonumber\\
               &=& \frac{1}{2\pi} \int dp \,dn \;  P[-i \p_\sigma,-\p^2_l]\left. f(a+\frac{1}{2} \ln\sigma,n) e^{i\,p\,(l-n\,\sigma)} \right\vert_{\sigma=1}\nonumber\\
               &=& P[-i \p_\sigma,-\p^2_l] \frac{1}{\sigma}\left. f(a+\frac{1}{2} \ln\sigma,\frac{l}{\sigma})\right\vert_{\sigma=1} \qquad .
\end{eqnarray}
\subsection{The case $k\;=\;\frac{\hbar^2}{4}$ and Zagier type deformations}
We now consider the case of our closed star product $\star:=T^{\mbox{\tiny{\rm closed}}}(\star^0_\hbar)$
together with a particular derivation $D_\hbar\in\mathfrak{Der}(\star)$ defined as setting $k=\frac{\hbar^2}{4}$ in the
expression (\ref{lambdaF}) and then intertwining by $T^{\mbox{\tiny{\rm closed}}}$ following (\ref{DERIV}).
We then proceed as in the subsection \ref{SectSL}, but use the following variables rather than (\ref{UVqr}):
\begin{equation}\label{UVqrbis}
\xi = q^{\frac{1}{2}}\frac{2}{\hbar}\sinh(\frac{\hbar}{2}\zeta) \quad, \quad \eta =
q^{\frac{1}{2}}\cosh(\frac{\hbar}{2}\zeta)  \quad
\end{equation}
Note that in contrast with the coordinates (\ref{UVqr}), the above one are not singular in the limit $\hbar\to 0$. Indeed, they become $\xi_0=q^{\frac{1}{2}}\zeta$ and $\eta_0=q^{\frac{1}{2}}$ in the limit.

\noindent Within these normalisations, the equation (\ref{wave})  then becomes:
\begin{equation}\label{wave-bis}
\partial_\xi\partial_\eta G\;=\;4k'\,\eta\xi\,G\;;
\end{equation}
with, as before,
\begin{equation}
G(q,r)  = q^{1/2} (1+ r^2\hbar^2)^{1/4}\,F(q,r)\;.
\end{equation}
Note that from (\ref{uF}), the classical  limit $\hbar\to0$ of the latter is prescribed to equal
\begin{equation}
F_{\mbox{\rm cl}}\;:=\;\delta_0(a)\otimes{\bf 1}_r\;.
\end{equation}
One way to achieve this requirement is to allow $k'$ to depend on $\hbar$ in such a way that the condition (\ref{kmu}) remains satisfied
\begin{equation}
\lim_{\hbar\to0}k'(\hbar)\;=\;0\;.
\end{equation}
More naturally, this corresponds to considering the $\hbar$-independent wave equation in the case $k'=0$:
\begin{equation}\label{wave-ter}
\partial_\xi\partial_\eta G\;=\;0\;,
\end{equation}
which trivially admits the $\xi$-independent solution:
\begin{equation}
G\;=\;G(\eta)\;=\:G\left(e^a\frac{\sqrt2}{2}\left(1+\sqrt{1+\hbar^2r^2}\right)^{\frac{1}{2}}\right)\;;
\end{equation}
with
\begin{equation}
G(e^a)\;:=\;e^{a}\,\delta_0(a)\;.
\end{equation}
The above discussion leads us to the particular solution:
\begin{equation}
u_\hbar(a,\ell)\;:=\;\frac{\sqrt2e^{2a}}{4\pi}\int_{-\infty}^\infty\,e^{ir\ell e^{2a}}\left(\frac{1+\sqrt{1+\hbar^2r^2}}{\sqrt{1+\hbar^2r^2}}\right)^{\frac{1}{2}}\,\delta_0\left(\,a\,+\,\frac{1}{2}\log
(\frac{1}{2}(1+\sqrt{1+\hbar^2r^2})\,)\right)\,{\rm d}r\;;
\end{equation}
that admits the correct limit:
\begin{equation}
u_0(a,\ell)\;=\;\delta_{e}(a,\ell)\;.
\end{equation}
The associated convolution operator $U$ produces a deformation
quantization that is invariant under the infinitesimal action of ${\mathfrak{sl}}_2(\R)$ on an open $\SS$-orbit
in $\R^2$ (the group $\SS$ acting by special linear transformations). The corresponding underlying geometry 
here being flat ($k'=0$). In particular, this class of solutions reproduces star products of the same type as the one considered by Connes and Moscovici in \cite{CM}, primarily constructed by Zagier as a $\mathfrak(sl)_2(\R)$-invariant deformation of the algebra of modular forms \cite{Za}.

\subsection{Unterberger type solutions and Weinstein's asymptotics}\label{Unterberger}
In \cite{Unt,Unt2} an associative $\SL$ invariant composition law, though not a star product, has been derived  in a totally different context by A.
and J. Unterberger, for the composition of symbols in the so-called Bessel calculus. We are going to show that a slight modification of their formula yields one of the simplest products in the family we have found.

\noindent As a prealable, let us notice, that in their works, these authors also make use of an interwiner operator linking an $\SL$- invariant composition law with an $AN$-invariant star product, but not the strongly closed one we have adopted. Instead of this one, the star product they use may be expressed in our framework as the one  obtained from (\ref{OpT}) with special weight function  ${\cal P}(\hbar \xi)=\cosh (\hbar \xi)$, as proven in \cite{BielMassar2}. 
\par\noindent The
convolution kernel of an intertwiner between an $\SL$-invariant star product $\#$ and a general
$AN\,$-invariant product, itself built from the Moyal star product twisted by an operator $T_{\cal P}$ is obtained from the operator $U$ we constructed in section \ref{SectSL} by:
\begin{equation}\label{fromone}
 U_{\cal P} = U  \circ (T^{{\mbox{\tiny{\rm closed}}}})^{-1} \circ T_{\cal P} \quad .
 \end{equation}
In particular, if $T_{\cal P}$ is of the form of eq. (\ref{OpT}), the kernel of $U_{\cal P}$ is then given as a superposition of modes,
\begin{equation}\label{ugenkernel}
 u_{\cal P} (a,l) = \int \, \phi(s') u^{\cal P}_{\hbar,s'}(a,l) \, ds' \quad , 
\end{equation}
with 
 \begin{equation}\label{uGenP}
u^{\cal P}_{\hbar,s}(a,l) =  e^{2a} \, h_s( e^{2a})\, \int \,
(1+\hbar^2 r^2)^{-\frac{1}{2}}\,{\cal P}({\rm arcsinh}(\hbar r))\, e^{ i s \sqrt{1+
\hbar^2 r^2} e^{2a}} \, e^{i r l e^{2a}} \, dr \quad,
\end{equation}
where $h_s(x)\propto  J_{
{\sqrt{k}}/{\hbar}}[\sqrt{s^2+ {k} /{\hbar^2}}\,\,
x] $.
 
\noindent An analog computation yields the kernel of the inverse operator $U^{-1}_{\cal P}$ 
again as a superposition of modes, each of them being given by 
\begin{equation} 
v^{\cal P}_{\hbar,s}(a,l) = e^{-2a} \, h_s( e^{-2a})\, \int
\, {\cal P}^{-1}({\rm arcsinh}(\hbar\,r)) e^{ i s \sqrt{1+\hbar^2
r^2} e^{-2a}} \, e^{-i r l} \, dr \quad .
\end{equation}

\noindent Let us plug  $B=0$ in \re{hs}, and normalize the modes accordingly to eqs (\ref{Cnorm}, \ref{Dnorm}) with $k=k'$ (see eqs (\ref{kmu}, \ref{kmu2}). The operator $U$ considered in  \cite{Unt,Unt2} then  corresponds to the value $s=0$  in \re{uGen}. This obviously amounts to pick up $\phi(s') = \delta (s')$ in eq. (\ref{ugenkernel}). By taking ${\cal P}(\hbar\,\xi)=\cosh (\hbar\,\xi)$, and denoting by $U_U$ the corresponding operator, one finds
\begin{equation}
 (U_{U} f) (a,l) = \frac{\sqrt{k}}{ \hbar} \int dx \; J_{\frac{\sqrt{k}}{\hbar}} ( {\sqrt{k}}\,x/\hbar) \, f(a-\frac{1}{2} \ln x,l\,x ).
\end{equation}
The inverse operator, $U^{-1}_U$ corresponds to the kernel resulting from the superposition
\begin{equation}
  v_{\hbar,s} (a,l) = \int \psi(s) \Psi_{\hbar,s}^{U}(a,l) ds
\end{equation}
of modes
\begin{equation}
 \Psi_{\hbar,s}^{U}(a,l) =\frac{\sqrt{k}}{ 2\,\pi\,\hbar}\left(\frac{\sqrt{k'}-i\,\hbar\,s(\hbar)}{\sqrt{k'}+i\,\hbar\,s(\hbar)}\right)^\frac {\sqrt{k}} {2\,\hbar}\, J_{\frac{\sqrt{k}}{\hbar}}(\sqrt{s^2 + k'/\hbar^2} \,x) \int    
 \frac{ e^{i \,s\, x \sqrt{1+\hbar^2\, r^2}} \, e^{- i r l}}{\sqrt{ (1+\hbar^2 \,r^2)}}\, dr 
 \end{equation}
that is obtained by using 
\begin{equation}
\psi(s)=(i-s)\,\delta'(s)
\end{equation}
in the superposition of modes like those appearing in eq. (\ref{Vsf}), normalized according to eq. (\ref{Dnorm}).
Explicitly,   one gets
\begin{equation}
 (U^{-1}_U f)(a,l) = \frac{\sqrt{k}}{ \hbar} \int dx \; J_{\frac{\sqrt{k}}{\hbar}} ({\sqrt{k}}\,x/\hbar) \, f(a+\frac{1}{2} \ln x ,\frac{  l}{x}).
\end{equation}
Apart from the $\hbar$-dependent pre-factors, these operators are exactly the ones found in \cite{Unt}, see equations (4.16) and (4.7).

\noindent Of course, the asymptotic expansions allows to build $U^{-1}$ perturbatively, as an expansion in $\hbar$. For illustrative purpose let us mention, that for a fixed value of $s$ we obtain at fourth order: 
\begin{eqnarray}
U^{-1}_s&=&\left(V_r -\frac i{4\,\pi}\,\left[1- i\,\hbar^2 r\,(\frac14+ i\,r)\right]\,\partial_r\,V_r\right)_{r=-s}+ O(\hbar^5)\nonumber \\
&=&\int \left[(1+\frac {\hbar^2}{16\,\pi}+i\frac{\hbar^2}2\,r) \delta[r+s]+\frac i{4\,\pi}(1-i\,\frac{\hbar^2}4\,r+ \pi\,\hbar^2\,r^2)\delta'(r+s)\right]\,V_r\, dr + O(\hbar^5)\nonumber\\
&&
\end{eqnarray}
\noindent By a long but straightforward computation, the above explicit expressions of $U_U$ and $U_U^{-1}$ yield
(for $k=1$) the following integral formula for the invariant star product on the hyperbolic plane $\DD$:
\begin{equation*}
u\,\sharp_\hbar \,v\,(x)\;=\;\frac{1}{16 \pi^3 \hbar^4}\, \int_{\DD\times\DD}\,K_\hbar\left(S_{\mbox{\rm can}}(x,y,z)\right)\,u(y)\,v(z)\,{\rm d}y\,{\rm d}z\;;
\end{equation*}
where ${\rm d}y$ denotes the Liouville measure on $\DD$ and where 
\begin{equation}\label{noyauK}
K_\hbar (\varpi)\;:=\; \int_0^\infty \, s^2 J_{\frac{1}{\hbar}} (s/\hbar)\,
 e^{\frac{i}{\hbar} s \varpi} \,{\rm d}s\;.
\end{equation}

\noindent Using tabulated Laplace transforms, one first computes
\begin{equation}
{\mathcal F}_\m(p)=\int_0^\infty \, e^{-pt} J_\mu (t) dt =
\frac{1}{p^{\mu +1}}
\frac{[1+(1+\frac{1}{p^2})^{1/2}]^{-\mu}}{\sqrt{1+\frac{1}{p^2}}}
\quad ,
\end{equation}
with $p=\varepsilon - i \varpi$, $\varepsilon \ll$ being a small
real part necessary to ensure convergence.
Using the following relations
\begin{equation} 1 + \frac{1}{p^2}  = \left\{ \bearlll
  1- \varpi^{-2} &\;& {\rm if}\ \varpi^2 >1 ,  \\[.3em]
  (\varpi^{-2} -1) \,{e}^{i \,\rm{sign}(\varpi)\pi} &\;& {\rm if}\ \varpi^2 < 1 ,  \eear\right .
\end{equation}
and 
\begin{equation} 1 + \sqrt{1 + \frac{1}{p^2}}  = \left\{ \bearlll
 1+ (1- \varpi^{-2})^{1/2} \in \R &\;& {\rm if}\ \varpi^2 >1 ,  \\[.3em]
 1+  (\varpi^{-2} -1)^{1/2} \,{e}^{i \,\rm{sign}(\varpi)\pi/2} &\;& {\rm if}\ \varpi^2 < 1 \;,  \eear\right .
\end{equation}
one finally gets
\begin{equation}
 {\mathcal F}_\m(-i \varpi)  = \left\{ \bearlll
 \frac{1}{\sqrt{1-\varpi^2}}\,\, e^{i \, \frac{\mu}{2}
f_W(\varpi)} &\;& {\rm if}\ \varpi^2 < 1 ,  \\[.3em]
  |\varpi|^{-(\m+2)} (\varpi^2 - 1)^{-1/2} [1+ (1- \varpi^{-2})^{1/2}]^{-\m} \,{e}^{i\, \frac{\pi}{2} (\m + 1) \rm{sign}(\varpi)} &\;& {\rm if}\ \varpi^2 > 1 \;;  \eear\right .
\end{equation}
where $f_W$ is the odd function  (\ref{fW}) defined by $f_W(S_{\mbox{\rm can}}):=S_W$.
Note that, when $\varpi^2 < 1$, the phase of ${\mathcal F}_\m$ is precisely given by $f_W(\varpi)$, while when $\varpi^2 > 1$, it is pure phase. The kernel $K_\mu$ is now simply described  by
\beq \label{KMU}
  K_\mu (\varpi) = -\frac{d^2}{d\varpi^2}{\mathcal F}_\m(-i \varpi).
\eeq
\noindent The phase $S_\mu$ of the kernel $K_\m$ can then be determined from \re{KMU}. For $\varpi^2 <1$, one gets
\begin{equation}\label{PhaseK}
S_\mu(\varpi) = \mu \frac{f_W(\varpi)}{2} - \arctan [\frac{ 3
 \mu \sin f_W(\varpi)}{\mu^2 -4 + (\mu^2+2) \cos
 f_W(\varpi)}] \quad .
\end{equation}
\noindent In other words, for $\varpi^2 <1$, which corresponds to triples of points in $\DD$ for which Weinstein's $S_W$
is well defined,  the above kernel $K_\hbar$ can be expressed under the WKB oscillatory form. The expression
of the corresponding phase is then the following:
\begin{equation}
S\;= \; \frac{S_W}{2} - \hbar\arctan [\frac{ 3
\hbar\sin S_W}{ 1 -4\hbar^2 + (1+2\hbar^2) \cos
S_W}] \quad .
\end{equation}
In particular, one has the following asymptotics:
\begin{equation*}
S\;\sim\;\frac{S_W}{2} \;,
\end{equation*}
agreeing with  A. Weinstein's picture in \cite{WeinsteinTr}.

\section*{Acknowledgments}
P.B. and Ph.S. acknowledge partial support from the IISN-Belgium (convention 4.4511.06). 
 P.B. acknowledges partial support from the IAP grant `NOSY' delivered by the Belgian Federal Government. S.D. and Ph.S.  warmly thank the IHES for its hospitality and the exceptional working conditions provided to its guests.

\appendix 
\section{Complements}\label{AppenB}

\subsection{On $\DD$ and $\MM$}
{\bf 1 Group type structures.} To obtain the expressions (\ref{PSIM}) and (\ref{PSID}), we proceed as follows. Consider for  instance the hyperbolic plane $\DD=SL_2(\R)/SO(2)=:G/K$. Consider the following usual presentation of $\gg:=\mathfrak{sl}_2(\R)$ as
generated over $\R$ by the elements $E,F,H$ with table $[E,F]=H\,,\,[H,E]=2E$ and $[H,F]=-2F$.
Then one may set $\kk:=\R.(E-F)$, $\la:=\R.H\subset\pp=:\kk^{\perp}$ and $\ln:=\R.E$. The group
$\SS$ may then be realized as the connected Lie subgroup of $G$ admitting $\ss:=\la\oplus\lN$ as Lie
algebra. Within these notations, the Iwasawa decomposition $G=\SS K$ yields the above-mentioned 
identification $\SS=\DD$ and the following global coordinate system:
\begin{equation}
\ss\longrightarrow\SS:(a,\ell):=aH+\ell E\mapsto\exp(a H)\exp(\ell E)\;.
\end{equation}
In this simple group context, the involution $\tilde{\sigma}$ of $G$ is nothing else than the Cartan involution
associated with the data of $K$. The symmetry at the base point $K$ of $G/K$ therefore reads
$s_K(gK):=\tilde{\sigma}(g)K$ which for ${\bf a}\in A:=\exp(\la)$ and ${\bf n}\in N:=\exp(\lN)$ corresponds to $\tilde{\sigma}({\bf an})K={\bf a}^{-1}\tilde{\sigma}({\bf n})K$. Observing that $\sigma(E)=-F$, a small
computation then yields $\exp(\ell F)\exp(-\frac{1}{2}\log(1+\ell^2)H)\exp(-\ell E)\in K$ hence  the above expression of $\Psi^\DD$. The case of the Poincar\'e orbit is similar, details can be found in \cite{PierreStrict}.

\subsection{Co-adjoint orbits of $iso(1,1)$: the Poincar\'e plane}

Let us emphasize that when the Lie group under consideration is not semi-simple (e.g. the Poincar\'e group $Iso(1,1)$, in opposition to the $\SL$ group), only the co-adjoint orbits make sense {\it a priori\/} in the framework of subsection (\ref{DefandProp}). An elementary calculation shows that on $Iso(1,1)$ the co-adjoint orbits consist generically into hyperbolic cylinder  sheets, otherwise  remain four  planes and a line of fixed point. If we denotes by $b$, $e_0$ and $e_1$ a basis of  generators of   $iso(1,1)$ that obey the commutation relations:
\begin{equation}
[b,e_0]=e_1\qquad,\qquad[b,e_1]=e_0\qquad,\qquad[e_0,e_1]=0\qquad,
\end{equation}
and by $\beta$, $\varepsilon_0$ and $\varepsilon_1$ a basis of the dual space $iso^*(1,1)$, the co-adjoint orbits are
\begin{eqnarray}
&&\mbox{\rm the generic ones:}\quad x=k\left\{\begin{array}l v\,\beta +  \cosh (\alpha)\, \varepsilon_0 -  \sinh(\alpha)\, \varepsilon_1\ ,\\
v\,\beta -   \sinh (\alpha)\, \varepsilon_0 +  \cosh(\alpha)\, \varepsilon_1 
\end{array}\right.
\quad k\neq 0,\quad v\in \R,\ a \in \R\\
&&\mbox{\rm the four null ones:}\quad x= v\,\beta \pm   \alpha( \varepsilon_0 \pm\varepsilon_1)\quad,\quad  v\in \R,\ \alpha \in \R_0^+\\
&&\mbox{\rm the pointlike ones:}\quad x= v\beta \quad ,\quad  v=C^{te}\in \R. 
\end{eqnarray}
The first ones are coset of the Poincar\'e group by the subgroup of translations in time (or in space); the second ones are coset obtained by dividing by light-like translation; both  are topologically $\R^2$,  $\alpha$ and $v$ providing coordinates on them.
The fundamental vector fields, on a generic orbit, are given by (denoting a point $x$ by its coordinates $\alpha$ and $v$):
\begin{eqnarray}
&&b^*_{( \alpha,v)}=\partial_\alpha\qquad,\qquad {e_0 ^*}_{( \alpha,v)}= - \sinh\alpha\,\partial_l\qquad,\qquad {e_1 ^*}_{( \alpha,v)}=  \cosh\alpha\,\partial_l\qquad,\\
&&b^*_{( \alpha,v)}=\partial_\alpha\qquad,\qquad {e_0 ^*}_{( \alpha,v)}=  \cosh\alpha\,\partial_v\qquad,\qquad {e_1 ^*}_{( \alpha,v)}=  -\sinh\alpha\,\partial_v\qquad,
\end{eqnarray}
and the symplectic form is given by:
\begin{equation}
\omega_{( \alpha,v)}=d\alpha\wedge dv\qquad.
\end{equation}
In terms of the $(a,l)=(\alpha/2,v)$-coordinates  \re{ANmult} used throughout this paper, these fundamental vector fields would be re-expressed as $H^* = -\p_a$, $E^*=-e^{-2a} \p_l=- ({e_0 ^*}_{(a,l)}+{e_1 ^*}_{(a,l)})$ and $F^*= -k\,e^{2a} \p_l=\pm k ({e_0 ^*}_{(a,l)}-{e_1 ^*}_{(a,l)})$.

An affine, torsion free, connection will be $Iso(1,1)$ invariant if its coefficients verify the two sets of equations:
\begin{eqnarray}
&&\Gamma^{\alpha}_{\beta\gamma}=\Gamma^{\alpha}_{\gamma\beta}\\
&&\xi^\mu \partial_\mu \Gamma^{\alpha}_{\beta\gamma}-\partial_{\mu}\xi ^\alpha \Gamma^{\mu}_{\beta\gamma}+\partial_{\beta}\xi ^\mu \Gamma^{\alpha}_{\mu\gamma}+\partial_{\gamma}\xi ^\mu \Gamma^{\alpha}_{\beta\mu}+\partial_{\beta\gamma}\xi ^\alpha=0\qquad,\qquad \xi=\beta,\ \varepsilon_0,\ \varepsilon_1
\end{eqnarray}
These equations imply that in $(v,\alpha)$ coordinates,  only two connection coefficients are non vanishing, depending  on two constants $A$ and $B$:
\begin{equation}
\Gamma^v_{\alpha,\alpha}= -v+A\qquad,\qquad\Gamma^\alpha_{\alpha,\alpha}=B 
\end{equation}
If we moreover require the connection to be symplectic ($\nabla \omega =0$), this implies that only $\Gamma^v_{\alpha,\alpha}= -v+A$ is non zero. 

Finally imposing that the symmetry transformations (\ref{symPpl}) preserve the connection imposes that $A=0$.
On a symplectic manifold $(M,\omega,s)$ such a connection: torsionless, preserving the two-form $\omega$, and invariant with respect to the symmetries $s$ is unique.It constitute the so-called Loos connection (\cite{Bithese}), intrinsically  defined by
\begin{equation}
\omega_x(\nabla_{X}Y,Z)=\frac 12 X_x\omega_x(Y+s_{x\star}Y,Z)
\end{equation}
or in terms of the symmetry expressed in coordinates $x^\rho$ as
$x^\rho[s_P(Q)]=\text{\rm s}^\rho(x^\mu[P],x^\nu[Q])$:

$$\Gamma^\rho_{\sigma\tau}(x^\mu[P])=\left.\frac{\partial^2\text{\rm s}^\rho(x^\mu[P],x^\nu[Q])}{\partial x^\sigma[Q]\partial x^\tau[Q]}\right\vert_{Q=P}
$$
The geodesic differential equations are
\begin{equation}\ddot a =0\qquad,\qquad \ddot\ell-4\,\ell\,{\dot a} ^2=0
\end{equation}
from which we infer immediately the equations of the affine geodesic curves (in term of an affine parameter $s$, starting from the point of coordinates $(a_0, \ell_0)$ with tangent vector, in natural components, $(p_0,q_0 )$:
\begin{equation}
 a =p_0\, s+ a_0\qquad,\qquad \ell =\ell_0\cosh (2\,p_0\, s) + q_0\frac{\sinh (2\,p_0\, s)}{2\,p_0}\qquad.
\end{equation}
From these we may recover the symmetry (\ref{symPpl}) and mid-point (\ref{midPpl}) equations; but let us emphasize that these are defined directly in terms  of the group action (\ref{ANmult}) and the involution (\ref{PSIM}).

\addcontentsline{toc}{chapter}{\hbar mberline{}Index}
\addcontentsline{toc}{chapter}{\hbar mberline{}Bibliography}

\bibliography{master2}
\end{document}